\documentclass[sigconf]{acmart}

\usepackage{booktabs} 

\usepackage{siunitx}
\usepackage{multirow}
\usepackage{adjustbox}
\usepackage{color}
\usepackage[utf8]{inputenc}
\usepackage[T1]{fontenc}
\usepackage{xspace}
\usepackage{tabularx}
\usepackage{graphicx}

\usepackage{algorithm}
\usepackage{algorithmic}

\usepackage{enumitem}
\usepackage{balance}

\usepackage{subfig}

\DeclareMathOperator*{\argmin}{arg\,min}

\makeatletter
\DeclareRobustCommand\onedot{\futurelet\@let@token\@onedot}
\def\@onedot{\ifx\@let@token.\else.\null\fi\xspace}

\def\ie{\emph{i.e}\onedot}

\def\wrt{w.r.t\onedot} 

\makeatother

\makeatletter
\newcommand\footnoteref[1]{\protected@xdef\@thefnmark{\ref{#1}}\@footnotemark}
\makeatother

\setcopyright{acmlicensed}

\theoremstyle{definition}

\begin{document}

\copyrightyear{2019}
\acmYear{2019}
\acmConference[MM '19]{Proceedings of the 27th ACM International Conference on Multimedia}{October 21--25, 2019}{Nice, France}
\acmBooktitle{Proceedings of the 27th ACM International Conference on Multimedia (MM '19), October 21--25, 2019, Nice, France}
\acmPrice{15.00}
\acmDOI{10.1145/3343031.3351030}
\acmISBN{978-1-4503-6889-6/19/10}

\fancyhead{}

\title{Cross-Modal Subspace Learning with \\Scheduled Adaptive Margin Constraints~\textsuperscript{*}}

\thanks{* Please cite the ACM MM 2019 version of this paper.}

\author{David Semedo}
\affiliation{%
  \institution{NOVALINCS}
  \streetaddress{}
  \city{Universidade NOVA de Lisboa}
  \state{Portugal}
}
\email{df.semedo@campus.fct.unl.pt}

\author{João Magalhães}
\affiliation{%
  \institution{NOVALINCS}
  \streetaddress{}
  \city{Universidade NOVA de Lisboa}
  \country{Portugal}}
\email{jm.magalhaes@fct.unl.pt}

\renewcommand{\shortauthors}{D. Semedo et al.}

\begin{abstract}
Cross-modal embeddings, between textual and visual modalities, aim to organise multimodal instances by their semantic correlations. State-of-the-art approaches use maximum-margin methods, based on the hinge-loss, to enforce a constant margin $m$, to separate projections of multimodal instances from different categories.
In this paper, we propose a novel scheduled adaptive maximum-margin (SAM) formulation that infers triplet-specific constraints during training, therefore organising instances by adaptively enforcing inter-category and inter-modality correlations. This is supported by a scheduled adaptive margin function, that is smoothly activated, replacing a static margin by an adaptively \textit{inferred} one reflecting triplet-specific semantic correlations while accounting for the incremental learning behaviour of neural networks to enforce category cluster formation and enforcement.
Experiments on widely used datasets show that our model improved upon state-of-the-art approaches, by achieving a relative improvement of up to $\approx 12.5\%$ over the second best method, thus confirming the effectiveness of our scheduled adaptive margin formulation.
\end{abstract}

\ccsdesc[500]{Computing methodologies~Neural networks}
\ccsdesc[500]{Information systems~Multimedia and multimodal retrieval}

\keywords{Cross-modal embedding; Adaptive maximum-margin; neural networks; multimedia retrieval}

\maketitle

\graphicspath{{figs/}}

\section{Introduction}
Documents with both visual and textual data have very rich information that span across the two modalities. These modalities naturally co-occur, each adding a unique semantic perspective to a document instance.
In this paper, we address the task of cross-modal retrieval, in which one is interested in being able to, given one modality (e.g. text), search by relevant content from the other modality (e.g. images), and vice-versa, in an unified manner.
The field of cross-modal embedding learning, has been actively researched~\cite{Rasiwasia:2010:NAC:1873951.1873987, Feng:2014:CRC:2647868.2654902, 7298966,Fan:2017:CRL:3123266.3123369,8013822,Wang:2017:ACR:3123266.3123326,7346492,Xu:2018:MSL:3206025.3206033,DBLP:journals/corr/abs-1708-04308}, with the most widely used approach being representation learning, through subspace learning. The rationale is to solve the heterogeneity problem by learning a common space in which semantically equivalent instances will be structured close together. Namely, projections are learned for each modality, mapping original representation vectors to a semantically correlated space.
The maximum-margin formulation, which consists of a variant of the \emph{hinge loss}, has been adopted lately by most state-of-the-art approaches~\cite{Wang:2017:ACR:3123266.3123326}.
This loss function enforces a set of hinge loss constraints, over sampled triplets (\textit{target instance}; \textit{positive instance}; \textit{negative instance}). Namely, it enforces image and text instances of the same category to be close, and instances of different categories to be far apart by at least a fixed margin $m$.
These correlations are then grounded in statistical~\cite{Rasiwasia:2010:NAC:1873951.1873987, 7298966, 7780910}, semantic~\cite{Wu:2018:LSS:3240508.3240521,Wang:2017:ACR:3123266.3123326,Peng:2016:CSR:3061053.3061157,8013822,Xu:2018:MSL:3206025.3206033} and temporal correlations~\cite{Semedo:2018:TCR:3240508.3240665}.

\begin{figure}[t]
    \centering
    \includegraphics[width=\linewidth]{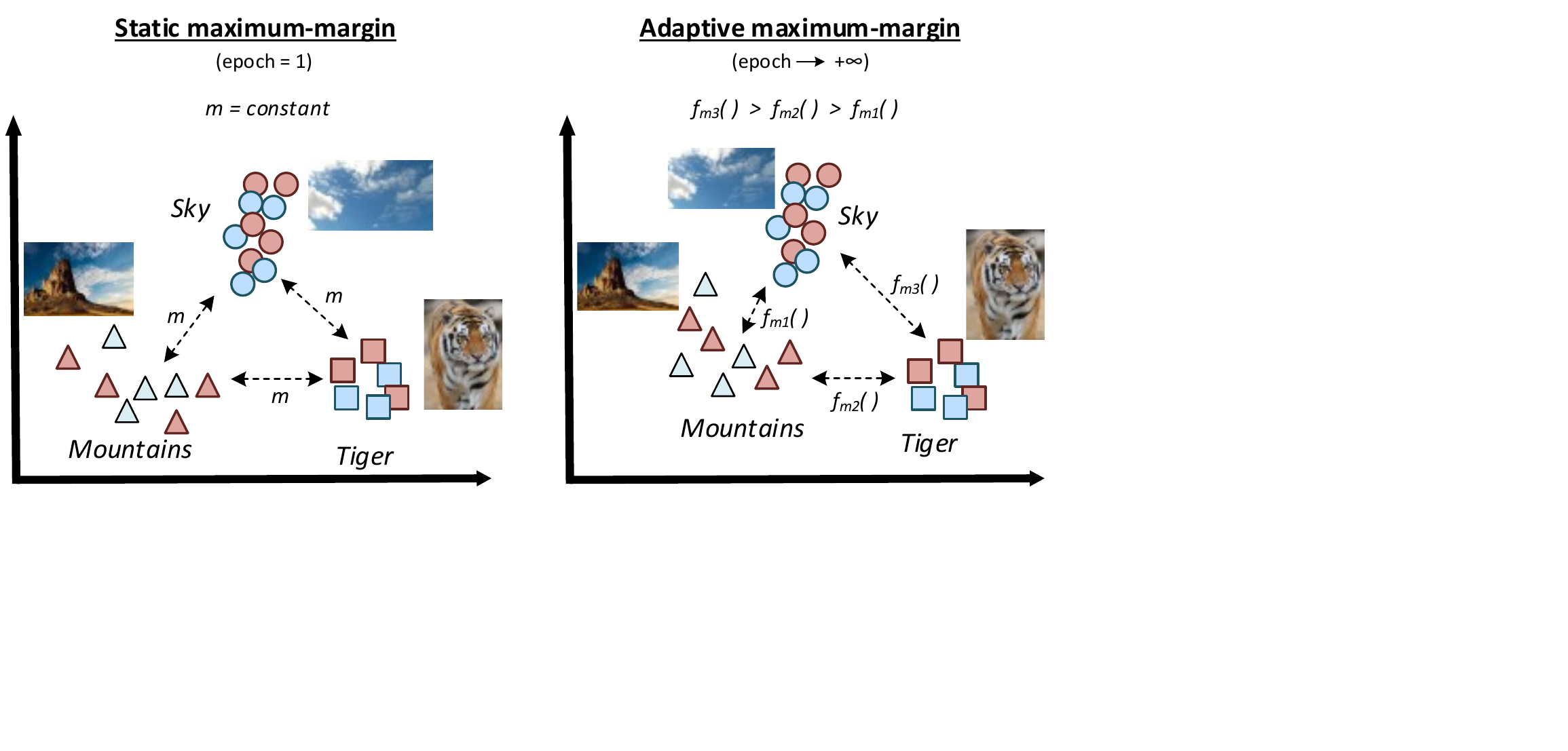}
\caption{Adaptive margin constraints are scheduled to be progressively activated during the training phase.}
  \label{fig:instance_structuring}
  \vspace{-10pt}
\end{figure}

In this paper we propose an adaptive neural structuring cross-modal subspace learning model (SAM), that dynamically organises instances on the new subspace according to their semantic similarity and inter-category correlations.
In particular the two main novelties of the proposed method are:
\begin{itemize}
    \item \textbf{Adaptive margin constraints:} we part ways with state-of-the-art methods based on the \emph{hinge-loss} function with a constant margin $m$ between different categories, and introduce a novel adaptive margin function $f_m(\cdot)$ that infers the margin constraints during training.
    \item \textbf{Scheduled activation of adaptive margins:} by considering the incremental learning behaviour of neural architectures \cite{Goodfellow:2016:DL:3086952}, we propose a novel \textit{scheduled learning algorithm} that progressively increases the parameters' degrees of freedom to allow a shift from coarse-grain (fixed margin $m$) to fine-grain (adaptive margins $f_m(\cdot)$) training, as the model converges to a stable solution. Figure~\ref{fig:instance_structuring} illustrates this shift from \textit{epoch 1} to \textit{epoch t}.
\end{itemize}

These contributions stem from the fact that the hinge-loss function does not adapt the constraints imposed by looking at the current subspace organisation, (e.g. clusters formed), at each training epoch $t$. We posit that semantic information used for subspace structuring should be directly incorporated in the ranking loss formulation, instead of adding extra terms to the main loss function. At the same time, the loss function should adapt the constraints imposed, at each training epoch, according to the current subspace structure and enforce semantic clusters formation, \ie promote grouping of instances of the same semantic category.

In summary, we formulate an adaptive maximum-margin model, which dynamically adapts subspace structuring constraints over triplets, by jointly using semantic similarity and subspace category clusters enforcement rules to obtain an effective semantic subspace organisation. Experiments on three cross-modal retrieval benchmark datasets, where we compare our method with a considerable number of existing methods, reveal that our model is highly effective, outperforming state-of-the-art works.

\section{Related work}

\noindent
\textbf{Cross-modal subspace learning. }
Learning cross-modal embeddings, between visual and textual data, has been an active research topic~\cite{Rasiwasia:2010:NAC:1873951.1873987, Feng:2014:CRC:2647868.2654902, 7298966,Fan:2017:CRL:3123266.3123369,8013822,Wang:2017:ACR:3123266.3123326}. In a pioneering work~\cite{Rasiwasia:2010:NAC:1873951.1873987}, Canonical Correlation Analysis~\cite{hotelling36cca} (CCA) was used to learn \textit{linear} projections for each modality, by learning a set of canonical coefficients, that define a subspace where modalities are maximally correlated. This approach was extended for the multi-label scenario, by using label information to establish correspondences between instances~\cite{7410823}. A multi-view kernel CCA formulation is proposed in~\cite{Gong:2014:MES:2584252.2584265}, where a joint space for visual, textual and semantic information is learned.

Lately, neural methods have proved to be highly effective at learning non-linear projections that capture complex non-linear correlations. The loss function definition is generally the core component, for which several variants have been proposed.
Deep Canonical Correlation Analysis (DCCA) was adopted in~\cite{7298966} to match images and text, using non-linear projections. DCCA is a non-linear version of CCA that exploits the fact that the CCA objective function can be formulated based on a matrix trace norm, allowing for gradient-based optimisation.
In~\cite{Feng:2014:CRC:2647868.2654902} a neural architecture, the Correlation Autoencoder (Corr-AE), with two uni-modal autoencoders (one per modality) is used,  enforcing correlation between learned hidden-representations. Instead of solely focusing in pairwise \emph{visual-textual} correlations, in~\cite{Peng:2016:CSR:3061053.3061157} extra constraints are added over inter-modal sample relations. Recently, in~\cite{8013822} authors model intra and inter-modality correlations, to unveil complex and fine-grain modality interactions.
An adversarial approach is proposed in~\cite{Wang:2017:ACR:3123266.3123326}, where a common subspace is learned by a mini-max game between a feature projector and a modality classifier.
An effective approach, common to several state-of-the-art approaches is \textit{triplet ranking loss}~\cite{2015arXiv150303832S}, in which different triplet mining strategies may be devised.

\vspace{3mm}
\noindent
\textbf{Subspace structuring constraints.}
Apart from maximising correlation between different modalities, additional constraints are usually added to the global loss function. In~\cite{pmlr-v77-kang17a} center-loss is used  to minimise intra-category invariance, under a metric learning approach. A successful approach consists of combining intra-modality semantic category and inter-modality pairwise similarity constraints~\cite{7780910,8013822,Wang:2017:ACR:3123266.3123326}. Such constraints are commonly enforced over sampled triplets. In~\cite{7780910} structure-preserving (hinge loss based) constraints with fixed margins, are used to push semantically similar instances closer to each other. In this paper, we follow a similar intuition, but instead we adaptively change the margin during training to enforce a per-category cluster formation and preservation.

\vspace{3mm}
\noindent
\textbf{Maximum-margin learning.}
To organise data by their semantic correlations, ranking loss is a widely adopted approach for cross-modal subspace learning due to its effectiveness~\cite{8019528,Wang:2017:ACR:3123266.3123326}. A set of similarity constraints are formulated under the hinge-loss, enforcing the similarity of positive instances to be far apart from similarity of negative ones, by at least a margin $m$. In state-of-the-art works, this margin is fixed with a constant value for all categories. In fact, this corresponds to a relaxation of the subspace structuring problem, in which the embedding's semantic similarities are neglected, thus possibly sacrificing optimal data organisation. Following this line of reasoning, \citet{DBLP:conf/iccv/LiZC15} replace the margin by the mean per joint error function, and in~\cite{8019528} the margin is replaced by the correlation of categories in the original feature space. We depart from the above methods by proposing an adaptive maximum-margin formulation that infers margin values during training.


\begin{figure*}[t]
    \centering
    \includegraphics[width=\linewidth]{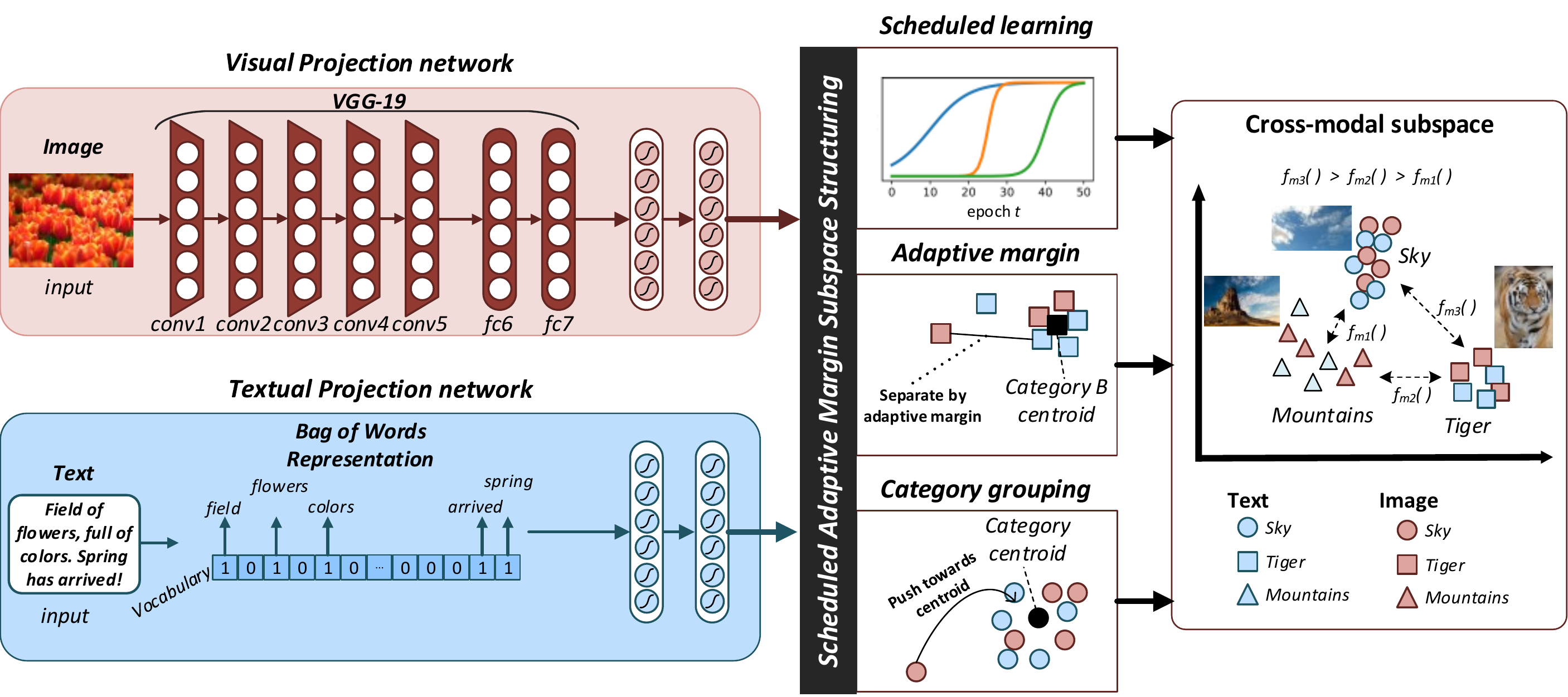}
\caption{SAM model architecture. The model is composed by two sub-networks coupled by the loss function $\mathcal{L}_{SAM}$. At each learning epoch $t$ the loss $\mathcal{L}_{SAM}$ imposes triplet-specific constraints, enforcing cluster formation/preservation and organising instances according to their semantic similarity.  }
  \label{fig:architecture}
\end{figure*}

\section{Cross-modal subspace structuring}

\subsection{Definitions}

Let $\mathcal{C}$ be a corpus of multimodal instances, where without loss of generality, the visual (images) and textual modalities are considered. Each instance $d^i \in \mathcal{C}$ is defined as $d^i=( x_V^i, x_T^i, l^i)$,
where $x_V^i \in \mathbb{R}^{D_V}$ and $x_T^i \in \mathbb{R}^{D_T}$ are the instance's image $d_V^i$ and text $d_T^i$ feature representations, respectively, and $l^i\in L$ the instance category. $L$ is the set of semantic categories.
Let $*\in\{V,T\}$ on the remainder of this paper, to avoid notation cluttering.

In cross-modal subspace learning, the goal is to learn a subspace in which instances' textual and visual elements, of the same semantic category, will be maximally correlated. The original feature spaces of $x_V$ and $x_T$ are dissimilar and cannot be used to perform cross retrieval, as they not only may have different dimensionalty but also encode different characteristics and semantics. To this end, for each original modality space, the goal is to learn the projections:
\begin{equation}
\mathcal{P}_{\theta_V}(\cdot): \mathbb{R}^{D_V} \mapsto \mathbb{R}^D \hspace{1cm} \mathcal{P}_{\theta_T}(\cdot): \mathbb{R}^{D_T} \mapsto \mathbb{R}^D
\end{equation}
mapping images $x_V$ and texts $x_T$ to a common cross-modal subspace, with dimensionality $D$. Similarity between two $\ell_2$ normalised projected sample modalities $x^i_*$ and $x^j_*$, is defined as \emph{cosine} similarity, and efficiently computed based on a dot product $s(x^i_*,x^j_*)=\mathcal{P}_{\theta_*}(x^i_*)\cdot\mathcal{P}_{\theta_*}(x^j_*)$, with the function $s(\cdot, \cdot)$, mapping to the range $[-1,1]$.

\subsection{Adaptive subspace learning}
Modality projections into cross-modal subspaces must capture both inter-category and inter-modality correlations in that subspace.
To this end, the cross-modal subspace learning problem is commonly formulated using a maximum-margin learning approach, by imposing a set of constraints over pairwise instance's similarity, on the target subspace~\cite{7410369,7780910,Wang:2017:ACR:3123266.3123326,Semedo:2018:TCR:3240508.3240665,8013822}.

For an anchor instance $x_*^a$, such constraints enforce the similarity of positive instances $s(x_*^a, x_*^p)$, i.e. sharing one category $l\in L$, to be higher than the similarity of negative samples $s(x_*^a, x_*^n)$, i.e. not sharing a category, by at least a margin $m$. This constraint is formulated as:
\vspace{-4pt}
\begin{equation}
    s(x_*^a, x_*^p) > s(x_*^a, x_*^n) + m.
\end{equation}
The constraint above is then enforced over each pair of instances, resulting in a considerable large set of constraints. For training, such constraints are then relaxed using the hinge loss~\cite{herbrich2000large}.

\subsubsection{Static maximum-margin formulation}
We start by formulating a loss function $\mathcal{L}$, under this framework, by imposing maximum-margin constraints over the two modality directions ($image\mapsto text$ and $text\mapsto image$), thus simultaneously capturing inter-modality and inter-category correlations. Namely, at every training epoch $t$, given triplets of the form $(x_*^a, x_*^p, x_*^n )$, where $x_*^p$ and $x_*^n$ stand for positive and negative instances, respectively, \wrt an anchor $x_*^a$, we compute the model loss,
\begin{equation}
\begin{split}
\mathcal{L}(t, \theta) &= \sum_{p,n} \underbrace{max(0, m-s(x_V^a, x_T^p) + s(x_V^a,x_T^n))}_{image\ \mapsto\  text} \ \ + \\
 &\ \ \ \ \sum_{p,n}  \underbrace{max(0, m-s(x_T^a, x_V^p) + s(x_T^a, x_V^n))}_{text\ \mapsto\  image},
    \label{eq:pairwise_loss_static_m}
\end{split}
\end{equation}
where $m$ denotes the margin and $\theta$ the model parameters. Note that unlike other cross-modal subspace learning works~\cite{8013822, Wang:2017:ACR:3123266.3123326, 8019528}, the positive instance $x_*^p$ from each triplet is \textit{only} the opposite modality of the same instance $d^i$, i.e. $x_V^p=x_V^a$ or $x_T^p=x_T^a$.  A negative sampling strategy is then applied to mine triplets that respect these conditions.

\subsubsection{Adaptive maximum-margin formulation}
The maximum-margin formulation defined in eq.~\ref{eq:pairwise_loss_static_m} assumes that \textit{any two instances from different categories are equally correlated}. This is reflected by the adoption of a constant margin $m$.

Inspired by maximum-margin structured SVM~\cite{Tsochantaridis:2005:LMM:1046920.1088722} formulation, we propose to (1) incorporate inter-category semantic correlations into the subspace structuring and (2) guide the projection learning algorithm, at each epoch, with structure preserving constraints that are derived from the current state of the subspace.  To achieve this, we propose an adaptive margin formulation, defined by a non-negative margin function $f_{m}(d^a, d^n, t)$, where $d^a$ and $d^n$ correspond to semantically different instances (i.e. belong to different categories) and $t$ denotes the current epoch of the subspace training algorithm.
The margin constraints, for every instance pair, at  epoch $t$, are then reformulated as:
\begin{equation}
    s(x_*^a, x_*^p) > s(x_*^a, x_*^n) + f_{m}(d^a, d^n, t).
\end{equation}
The rationale enclosed in this formulation is that for each pair of instances of different categories, $f_{m}(\cdot)$ outputs a margin that encodes the degree of separation between the considered categories. On every epoch $t$, the margin is linked to the pairwise correlation of the instances' original feature vectors and current subspace structure.
Accordingly, the adaptive subspace learning loss function $\mathcal{L}_{SAM}$, at epoch $t$ becomes:
\begin{equation}
\begin{split}
&\mathcal{L}_{SAM}(t, \theta) = \\
&\ \sum_{p,n} \underbrace{max(0, f_{m}(d^a, d^n, t)-s(x_V^a, x_T^p) + s(x_V^a,x_T^n))}_{image\  \mapsto\  text} \ \ + \\
&\ \sum_{p,n}  \underbrace{max(0, f_{m}(d^a, d^n, t)-s(x_T^a, x_V^p) + s(x_T^a, x_V^n))}_{text\ \mapsto\  image}.
    \label{eq:pairwise_loss_adaptive}
\end{split}
\end{equation}
Similarly to maximum margin methods, this formulation guides the model towards incorporating semantic information, regarding intra-category pairwise correlations.
However, we observe that the difference of the similarities between positive and negative instances are, on average, inter-category specific. Therefore, we account for the current subspace organisation (at each epoch $t$), to decide what should be the magnitude of the margin, \ie $f_m$.

\subsection{Scheduled activation of adaptive margins}
For neural subspace learning, in the first gradient updates, the subspace organisation is expected to be highly volatile, constantly changing at each epoch. It follows that for neural networks trained using stochastic gradient descent, it is not trivial to estimate beforehand when (i.e. at each epoch) is the model about to converge.
Thus, we propose an approximation strategy that imposes a hard (\ie a static high magnitude) margin on all triplet constraints on the first few epochs. This allows the model to find an initial coarse organisation of the subspace. Then, as the number of epochs progress, the static constraints give way to triplet specific constraints, that better capture the fine-grain interactions among instances.

\begin{figure}[h]
\vspace{-4pt}
    \centering
    \includegraphics[width=0.7\linewidth]{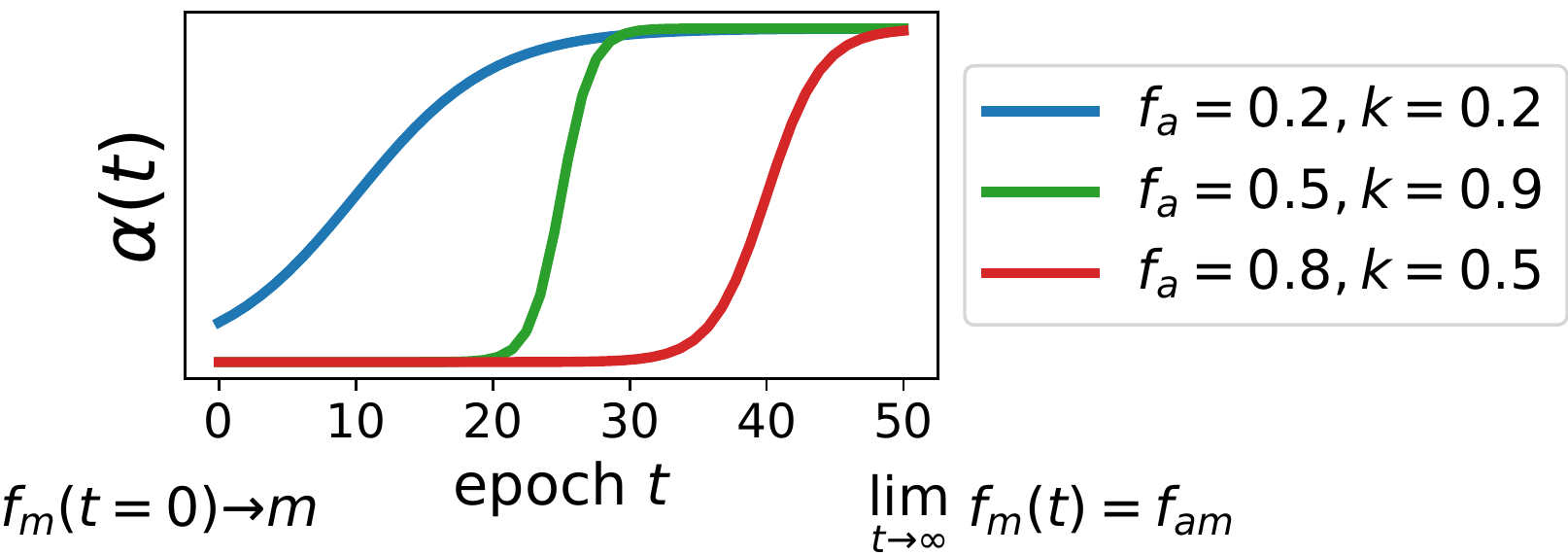}
\caption{Plot of $\alpha(t)$ with $n_e=50$. The scheduling training enables a smooth transition from static margins to adaptive margins. Best viewed in color.}
\label{fig:scheduling_function}
\end{figure}

Inspired by adaptive strategies for neural network training, such as the Adam~\cite{DBLP:journals/corr/KingmaB14} optimiser, which schedules different learning rates, we propose a smoothed scheduled shift function from static to an adaptive maximum-margin formulation, as the training algorithm converges (Figure~\ref{fig:scheduling_function}). To this end, a scheduled adaptive margin function $f_m$ is defined as:
\begin{equation}
\begin{gathered}
f_m(d^a, d^n, t) = \alpha(t)\cdot f_{am}(d^a, d^n, t) + (1-\alpha(t))\cdot m \\
\text{s.t. } \ \  \alpha(t)=\frac{1}{1+e^{-k\cdot(t-f_a\cdot n_e)}},
\label{eq:activation_factor}
\end{gathered}
\end{equation}
where the $\alpha(t)$ is a scheduler function, defined as a \emph{compressed} \textit{sigmoid}, that gradually activates the adaptive margin, according to the current epoch $t$.
The $\alpha(t)$ function is defined by a smoothing term $k$, the total number of epochs $n_e$ and an activation factor $f_a\in[0,1]$. Figure~\ref{fig:scheduling_function} illustrates how each parameter is used to define $\alpha(t)$.

\subsection{Adaptive margin function definition}
In this section we describe how the adaptive margin function $f_{am}(d^a, d^n, t)$ is materialised. We formulate $f_{am}$ such that it implements an adaptive margin, encoding: a) the semantic correlation -- estimated from original modality features -- between instances from different categories, and b) cluster formation enforcement, for each semantic category, according to the epoch $t$ of the algorithm. Figure~\ref{fig:architecture} illustrates $f_{am}$ components. In particular, we define the adaptive margin function as
\begin{equation}
    f_{am}(d^a, d^n, t) = \lambda\cdot f_{ms}(d^a, d^n) + (1-\lambda)\cdot f_{mc}(d^a, d^n,t),
\end{equation}
where $f_{ms}$ quantifies semantic correlation, and $f_{mc}$ the similarity between category clusters at epoch $t$, of two instances $d^a$ and $d^n$. The parameter $\lambda$ models the trade-off between the two components.

\vspace{3mm}
\noindent\textbf{Semantic inter-category pairwise correlations.}
From a semantic standpoint, pairwise correlations across categories, will be different (e.g. instances from category \emph{sky} are expected to be more correlated with instances from \emph{clouds} than from \emph{flowers}). In our neural subspace structuring model, the function $f_{ms}$ accounts for such semantic correlations by evaluating similarity on each modality original spaces. The function $f_{ms}$ is then defined as:
\begin{equation}
\begin{split}
f_{ms}(d^a, d^n) = \frac{|| x^i_V-x^n_V||_{2}+|| x^i_T-x^n_T||_{2}}{2}.
\end{split}
\label{eq:margin_semantic}
\end{equation}
From the definition, $f_{ms}$ averages the semantic similarity of both visual and textual modalities, extracted from the modalities' original feature space.
The output of this function is normalized to $[0,1]$.

\vspace{3mm}
\noindent\textbf{Category cluster formation and preservation.}
Given a randomly initialised neural network model, it can converge to different local optima, thus resulting in different subspace organisation. From this observation, we pose that for near convergence epochs, it is important to restrict model updates, preserving currently formed category clusters and forcing instances to move towards their category cluster.
As a generalization, the centroid of a given category $l$ is computed as:
\begin{equation}
\begin{split}
\mathcal{P}_{*c}(l, t)=\frac{1}{|\{x^j_*:l^j=l\}|}\cdot\sum_{x^k_*\in\{x^j_*:l^j=l\}}\mathcal{P}_{\theta_*}(x_*^k;t),
\end{split}
\label{eq:centroid_computing}
\end{equation}
To materialise the described behaviour, we rely on the cosine distance $d$ to define $f_{mc}$ as:
\begin{equation}
\begin{split}
&f_{mc}(d^a, d^n, t) =\\
&\frac{1}{2}\cdot \big[d(\mathcal{P}_{Vc}(l^a, t), \mathcal{P}_{Vc}(l^n, t)) + d(\mathcal{P}_{Tc}(l^a, t), \mathcal{P}_{Tc}(l^n, t))\big],
\end{split}
\label{eq:margin_centroid}
\end{equation}
where for a given category $l$, $\mathcal{P}_{Vc}(l, t)$ and $\mathcal{P}_{Tc}(l, t)$ denote the centroid of the visual and textual projections, at epoch $t$. $d$ stands for the cosine distance $1-s(\cdot, \cdot)$, with $s$ being normalised \textit{a priori} to $[0,1]$ range.
Essentially, given a pair of instances, $f_{mc}$ evaluates the distance between the corresponding category centroids, for both visual and textual projections.
Grounding the margin on $f_{mc}$ simultaneously enforces cluster formation and preservation. This is achieved because during training, the function $f_{mc}$ will simultaneously attempt to preserve the current subspace organisation and push bad aligned projections towards the corresponding category centroid.

\subsection{Neural architecture}
To learn projections $\mathcal{P}_{\theta_V}(\cdot)$ and\ $\mathcal{P}_{\theta_T}(\cdot)$, we consider a two decoupled network architecture, to learn non-linear mappings, predominant across multiple state-of-the-art works~\cite{Ngiam:2011:MDL:3104482.3104569,Fan:2017:CRL:3123266.3123369,Wang:2017:ACR:3123266.3123326,7298966, Feng:2014:CRC:2647868.2654902}. The networks are jointly trained by a common loss function $\mathcal{L}_{SAM}$. For each modality, a feedforward network $f_*$ maps original modality representations onto $\mathcal{S}$, comprising 2 fully connected layers (with dimensions 1024 and $D$, respectively)  and $tanh$ non-linearities. For semantically rich image feature representation,  each $x_V^i$ is obtained from a pre-trained CNN. Each $x_T^i$ is represented as a bag-of-words vector.  Then, $f_*$ takes its corresponding modality as input:  $d_V^i$ RGB image for $f_V$ and $d_T^i$ bag-of-words text representation for $f_T$. Figure~\ref{fig:architecture} depicts the full architecture.

\subsection{Training and inference}

We jointly learn both the cross-modal projections $\mathcal{P}_{\theta_V}(\cdot)$ and $\mathcal{P}_{\theta_T}(\cdot)$, while adaptively performing neural subspace structuring, by minimising the function:
\begin{equation}
\begin{gathered}
\argmin_{\theta_V, \theta_T} \mathcal{L}_{SAM}(\theta_V, \theta_T)
\label{eq:general_loss}
\end{gathered}
\end{equation}
where $\mathcal{L}_{SAM}$ adaptively  organizes instances according to their inter-category and inter-modal correlations.
Pseudocode is illustrated in algorithm~\ref{alg:algorithm}.
A stochastic sampling strategy is adopted, in which to evaluate $\mathcal{L}_{SAM}(\theta_V,\theta_T)$, negative samples are sampled directly from mini-batches. At each epoch, all samples are \emph{seen} by the network. This approach severely reduces the model complexity, while still achieving convergence. The whole model is then optimised using Stochastic Gradient Descent.

\renewcommand{\algorithmicrequire}{\textbf{Input:}}
\renewcommand{\algorithmicensure}{\textbf{Initialization:}}
\renewcommand{\algorithmicrepeat}{\textbf{repeat until convergence:}}

\newcommand{\INDSTATE}[1][1]{\STATE\hspace{#1\algorithmicindent}}
\newcommand{\INDFOR}[1]{\hspace{\algorithmicindent} \FOR{#1}}
\begin{algorithm}[t]
	\algsetup{linenosize=\scriptsize}
	\caption{Pseudocode for SAM optimisation.}
	\label{alg:algorithm}
	\begin{algorithmic}[1]
		\ENSURE  Corpus  $\mathcal{C}=\{d^1, \ldots, d^n\}$ of multimodal instances, with $d^i=(x^i_V, x^i_T, l^i)$;\\
		Initialise network weights: $\theta_V, \theta_T$;\\
		Hyperparameters: $\lambda$, $k$, $f_a$, subspace dimensionality $D$, learning rate $\eta$, mini-batch size $b$;

		\STATE \textbf{repeat until convergence:}
		\INDFOR{$t$ epochs}
		\INDSTATE{Sample mini-batch to create triplets of the form $(x_V^i, x_T^i, x_T^n)$ and $(x_T^i, x_V^i, x_V^n)$;}
		\INDSTATE{Update $\theta_V$ and $\theta_T$ through BP, with stochastic gradients, using $\alpha(t)$:}
		\INDSTATE{${ \theta_V  } \leftarrow { \theta_V  }-\eta \cdot { \nabla  }_{\theta_V}\frac { 1 }{ b } \left( \mathcal{L}_{SAM} \right)$;}
		\vspace{1pt}
		\INDSTATE{${ \theta_T  } \leftarrow { \theta_T  }-\eta \cdot { \nabla  }_{\theta_T}\frac { 1 }{ b } \left( \mathcal{L}_{SAM} \right)$;}
		\INDSTATE{Update the weight of the adaptive margin:}
		\INDSTATE{$\alpha(t+1)\leftarrow \frac{1}{1+e^{-k\cdot((t+1)-f_a\cdot n_e)}}$;}
		\ENDFOR
		\RETURN projection networks, $\mathcal{P}_{\theta_V}(\cdot)$ and $ \mathcal{P}_{\theta_T}(\cdot)$.
	\end{algorithmic}
\end{algorithm}

\section{Evaluation}

\subsection{Datasets}
We evaluate our proposed methods in three widely used cross-modal retrieval benchmark datasets.

\noindent\textbf{Wikipedia~\cite{Rasiwasia:2010:NAC:1873951.1873987}.} Comprised by a total of 2,866 \emph{visual-textual} pairs, extracted from Wikipedia's "featured articles", where each article is accompanied by a single image. Each article is annotated with 10 semantic categories. We split the dataset following~\cite{Rasiwasia:2010:NAC:1873951.1873987, Feng:2014:CRC:2647868.2654902,Peng:2016:CSR:3061053.3061157}, with 2,173 instances for training, 231 for validation, and 462 for testing.

\noindent\textbf{NUS-WIDE~\cite{nus-wide-civr09}}. The NUS-WIDE dataset is comprised by a total of 269,648 instances (images and corresponding tags), from the Flickr network, annotated with one or more categories from a total of 81 distinct semantic categories. For comparison, we follow the protocol of~\citet{8013822}: only instance pairs that belong to a single category are kept and the instances from the 10 categories with more instances\footnote{\label{note1}Top-10 categories: 'person', 'animal', 'sky', 'window', 'water', 'flowers', 'food', 'toy', 'grass', 'clouds'.} are chosen. This results in more than 60,000 instances. Splits are created following~\cite{8013822}, resulting in 23,661 instances for testing, 5,000 for validation and the remaining for training.\\
\noindent\textbf{NUS-WIDE-10K} is a subset of NUS-WIDE created by strictly following the protocol of~\cite{Feng:2014:CRC:2647868.2654902}: the 10 categories with more instances\footnoteref{note1} are chosen, and for each category, 1000 instances are sampled. Only pairs that belong to a single category are considered. Three splits, equally balanced \wrt to number of instances per category, are sampled randomly: 8,000 instances for training, 1,000 for validation and 1,000 for testing.

\noindent\textbf{Pascal Sentence~\cite{Rashtchian:2010:CIA:1866696.1866717}.} Comprised by 1,000  \emph{visual-textual} pairs, from the 2008 PASCAL development kit, categorised within 20 categories, with instances evenly distributed across categories. We follow~\cite{Feng:2014:CRC:2647868.2654902, Peng:2016:CSR:3061053.3061157} and randomly and evenly split the dataset with 800 instances for training, 100 for validation and 100 for testing.

\subsection{Methodology}
We evaluate the retrieval performance using mean Average Precision ($mAP$), which is the standard evaluation metric for cross-modal retrieval~\cite{Rasiwasia:2010:NAC:1873951.1873987, Wang2016ACS, Feng:2014:CRC:2647868.2654902,Wang:2017:ACR:3123266.3123326, 7298966, 8013822}. We follow~\cite{Rasiwasia:2010:NAC:1873951.1873987,7006724,6587747,8013822} and compute $mAP$ for \emph{all the  retrieved results}. For $mAP$, an instance is relevant if it has the same category. Two tasks are evaluated: 1) \emph{Image-to-Text} retrieval ($I\mapsto T$) and 2) \emph{Text-to-Image} ($T\mapsto I$) retrieval. Core parameters of SAM are analysed to assess their impact in the performance. Each $mAP$ result reported of our method corresponds to the average of 5 runs.

\begin{table*}[t]
\caption{mAP performance results across different datasets. The second half of the table concern deep-learning methods.}
\label{table:results}
\centering
\centering
\begin{tabularx}{\linewidth}{l *{9}{X}}
\toprule
\multirow{2}{*}{\textbf{Method}} & \multicolumn{3}{c}{\textbf{Pascal Sentences}} & \multicolumn{3}{c}{\textbf{NUS-WIDE-10k}} & \multicolumn{3}{c}{\textbf{Wikipedia}} \\
   & $I\mapsto T$ & $T\mapsto I$ & Avg & $I\mapsto T$ & $T\mapsto I$ & Avg & $I\mapsto T$ & $T\mapsto I$ & Avg \\
\hline
CCA~\cite{hotelling36cca} & 0.203  & 0.208  & 0.206 & 0.167 & 0.181 & 0.174 & 0.298 & 0.273 & 0.286\\
CFA~\cite{Li:2003:MCP:957013.957143} & 0.476  & 0.470  & 0.473 & 0.406 & 0.435 & 0.421& 0.319 & 0.316 & 0.318\\
KCCA~\cite{6788402} & 0.488  & 0.446  & 0.467 & 0.351 & 0.356 & 0.354 & 0.438 & 0.389 & 0.414\\
LGCFL~\cite{7006724} & 0.539  & 0.503  & 0.521 & 0.453  & 0.485  & 0.469 & 0.466 & 0.431 & 0.449\\
JRL~\cite{6587747} & 0.563  & 0.505  & 0.534 & 0.466  & 0.499  & 0.483 & 0.479 & 0.428 & 0.454 \\
\hline
Corr-AE~\cite{Feng:2014:CRC:2647868.2654902} & 0.532  & 0.521  & 0.527 & 0.441  & 0.494  & 0.468 & 0.442 & 0.429 & 0.436 \\
DCCA~\cite{7298966} & 0.568  & 0.509  & 0.539 & 0.452  & 0.465  & 0.459 & 0.445 & 0.399 & 0.422\\
CMDN~\cite{Peng:2016:CSR:3061053.3061157} & 0.544  & 0.526  & 0.535 & 0.492  & \underline{0.542}  & 0.517 & 0.487 & 0.427 & 0.457\\
Deep-SM~\cite{wei2016cross} & 0.560  & 0.539  & 0.550 & 0.497  & 0.478  & 0.488 & 0.478 & 0.422 & 0.450\\
ACMR~\cite{Wang:2017:ACR:3123266.3123326} & 0.538 & 0.544 & 0.541 & \underline{0.519} & \underline{0.542} & \underline{0.531} & 0.468 & 0.412 & 0.440\\
CCL~\cite{8013822} & \underline{0.576}  & \underline{0.561}  & \underline{0.569} & 0.481  & 0.520  & 0.501 & \underline{0.505} & \textbf{0.457} & \underline{0.481}\\
\hline
 SAM ($\alpha(t)=1,\ \lambda=1$) &  0.586  & 0.590  & 0.588 & 0.539  &0.559  & 0.549 & 0.406 & 0.382 & 0.394 \\
SAM & \textbf{0.637} & \textbf{0.643} & \textbf{0.640} & \textbf{0.563} & \textbf{0.594} & \textbf{0.579} & \textbf{0.518} & \textbf{0.457} &
\textbf{0.487}\\
\bottomrule
\end{tabularx}%
\end{table*}

\subsection{Implementation details}
Networks are jointly trained using Stochastic Gradient Descent, with $0.9$ Nesterov momentum, and a learning rate $\eta = \num{5e-3}$, with a decay of $\num{1e-6}$. The model with lowest validation error is kept. Mini-batch size is set to $200$ for all datasets, and the total number of epochs is set to $100$. The projections dimension is set to $D=200$ and the margin $m=1.0$.
For each neuron, \emph{tanh} non-linearities are applied. Dropout with $p=0.1$ is applied to the first hidden layer.
Images representations are obtained by feeding each individual image through a pre-trained VGG-19~\cite{DBLP:journals/corr/SimonyanZ14a} convolutional network, and extracting the output of the last fully connected layer (\textit{fc7}). For texts, we adopt a BoW representation, with 1000-D vocabulary size for NUS-WIDE-10k and Pascal Sentences, and 3000-D for wikipedia.

\subsection{Cross-modal retrieval}
We compare our proposed approach, SAM, with a total of 11 state-of-the-art works, on the task of cross-modal retrieval. Namely, we compare against CCA~\cite{hotelling36cca}, CFA~\cite{Li:2003:MCP:957013.957143}, KCCA~\cite{6788402}, Corr-AE~\cite{Feng:2014:CRC:2647868.2654902}, JRL~\cite{6587747}, LGCFL~\cite{7006724}, DCCA~\cite{7298966}, CMDN~\cite{Peng:2016:CSR:3061053.3061157}, Deep-SM~\cite{wei2016cross}, ACMR~\cite{Wang:2017:ACR:3123266.3123326} and CCL~\cite{8013822}.

\vspace{5pt}
\noindent\textbf{Pascal sentences dataset.}
Table~\ref{table:results} shows the results obtained. Our method outperforms all the compared methods, on both  $I\mapsto T$ and $T\mapsto I$ settings. Namely, SAM achieved a relative improvement of $\approx 12.5\%$, with respect to the second best performing method, CCL. CCL models intra-modality and inter-modality correlations through distinct constraints, using a strategy that balances both types of correlation constraints. These are then superseeded by a ranking loss function in which a static margin is used. Instead, SAM adopts an adaptive margin formulation, in which intra and inter modality correlations are directly modelled in a single constraint. The best result was achieved with $\lambda=0.25$, $f_a=0.4$ and $k=0.1$.
SAM started smoothly activating the adaptive margin at about half the training epochs, revealing preference for starting using $f_{am}$ sooner.
The semantic similarity component $f_{ms}$ plays an important role in organising the space. Notwithstanding, the component $f_{mc}$ has revealed to be the most important one ($75\%$), effectively guiding the subspace structuring.

\vspace{5pt}
\noindent\textbf{NUS-WIDE-10k dataset.}
From the results on table~\ref{table:results}, we can see that our method also achieved the best performance when compared to all methods, on both cross-modal retrieval directions. It outperformed both traditional cross-media models (top rows of table~\ref{table:results}) and the most recent deep learning methods. With respect to the second best performing method, ACMR, which uses an adversarial approach for subspace learning, we obtain a relative improvement of $\approx 9\%$, on the average of $T\mapsto I$ and $I\mapsto T$. This confirms the importance of moving  towards an adaptive margin formulation.  The best result was obtained with $\lambda=0.05$, $f_a=0.9$ and $k=0.1$. Hence, in contrast to the results on the Pascal sentences dataset, the method started activating the adaptive margin near the last epochs of training. Moreover, once again, more importance was given to the cluster enforcement and preservation ($95\%$ of the weight). Our method obtains a high $mAP$ on both directions, but performs better on the $T\mapsto I$ direction. We believe that the reason is that visually, some categories have very similar content (e.g. \emph{sky} vs. \emph{clouds}). However, the text in this dataset correspond to tags, which due to the sparsity of BoW representation, turns out to have good discriminative properties.

\vspace{5pt}
\noindent\textbf{Wikipedia dataset.} As with the previous datasets, our method outperforms  all the compared methods. On the Wikipedia dataset, categories are very broad (e.g. Art \& Architecture, Media, etc.), with texts and images of the same category being highly diverse. Therefore, in this dataset, given the small amount of instances available for training, it is harder to align modalities. As this is reflected in original feature representations, the function $f_{ms}$, which organises instances according to semantic similarity on original features, ends up not helping structuring the space. Supporting this observation is the fact that the best result was obtained with $\lambda=0.05$. The category cluster formation and preservation, enforced by function $f_{mc}$ provides the major contribution to the effectiveness.

To further complement our evaluation, we also compare our method against CMOLRS~\cite{8019528}, which formulated the margin as an original-feature driven margin that is fixed during training, i.e. using only a simplified version of $f_{ms}$ factor of SAM. On the Wikipedia dataset, CMOLRS achieved a mAP@100 of $0.413$ while SAM achieves a mAP@100 of $0.541$. As authors only report mAP@100, we did not included it in table~\ref{table:results}. This confirms the importance of dynamically adjusting margin values during training and of the novel cluster formation and preservation component $f_{mc}$.

\begin{figure*}[!t]
    \centering
    \begin{minipage}{0.35\textwidth}
    \centering
    \vspace{6mm}
    \includegraphics[width=0.9\linewidth]{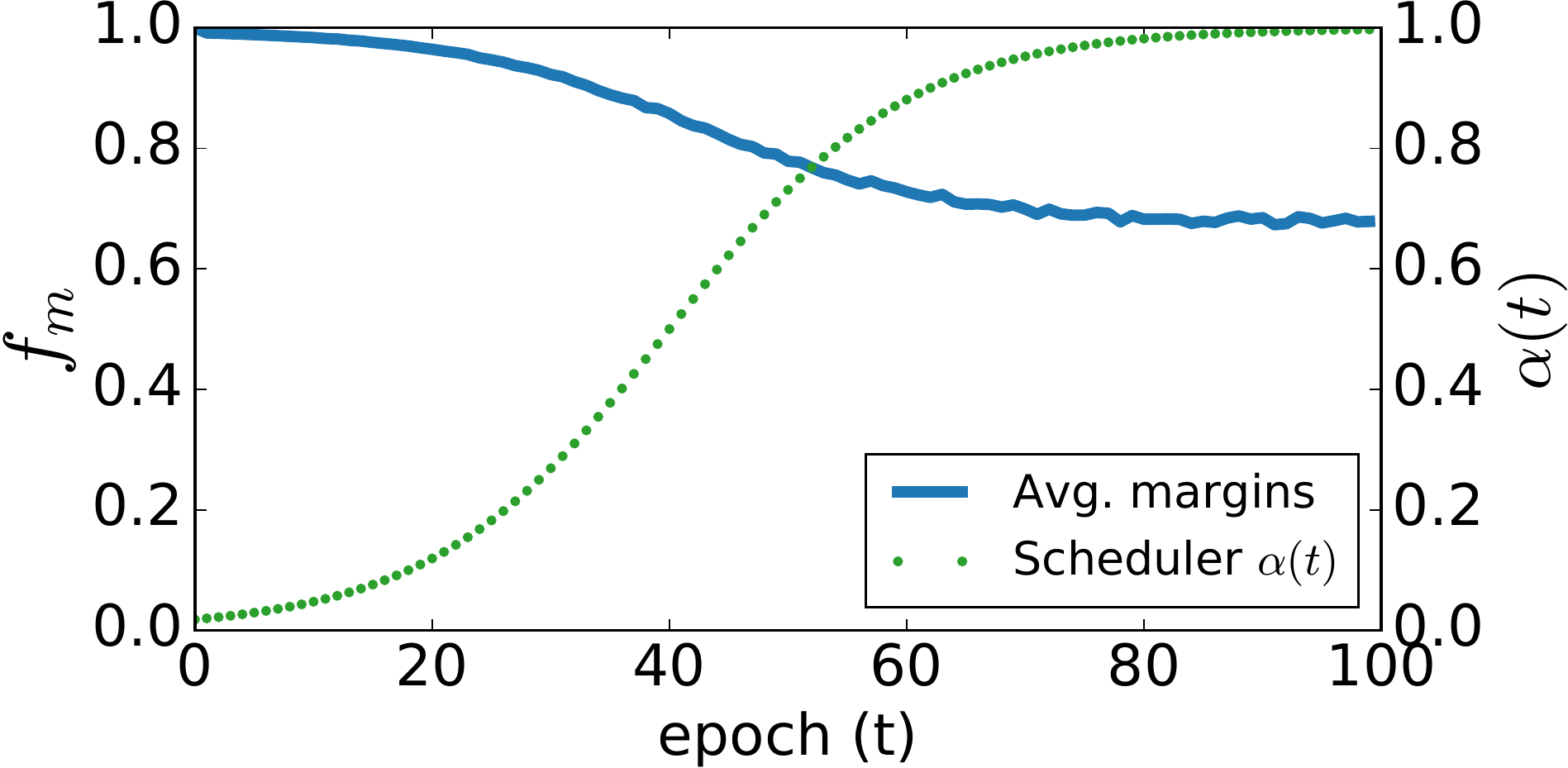}
    \caption{Global average adaptive margin $f_m$ over training epochs (t). The left y-axis corresponds to the $f_m$ value and the right y-axis to the scheduling function $\alpha(t)$ value. }
    \label{fig:margin_statistics_avg_epoch}
    \end{minipage}
    \qquad
    \begin{minipage}{0.55\textwidth}
    \centering
    \subfloat[Scheduled Adaptive Margins between 3 categories.]{{\includegraphics[width=0.47\linewidth]{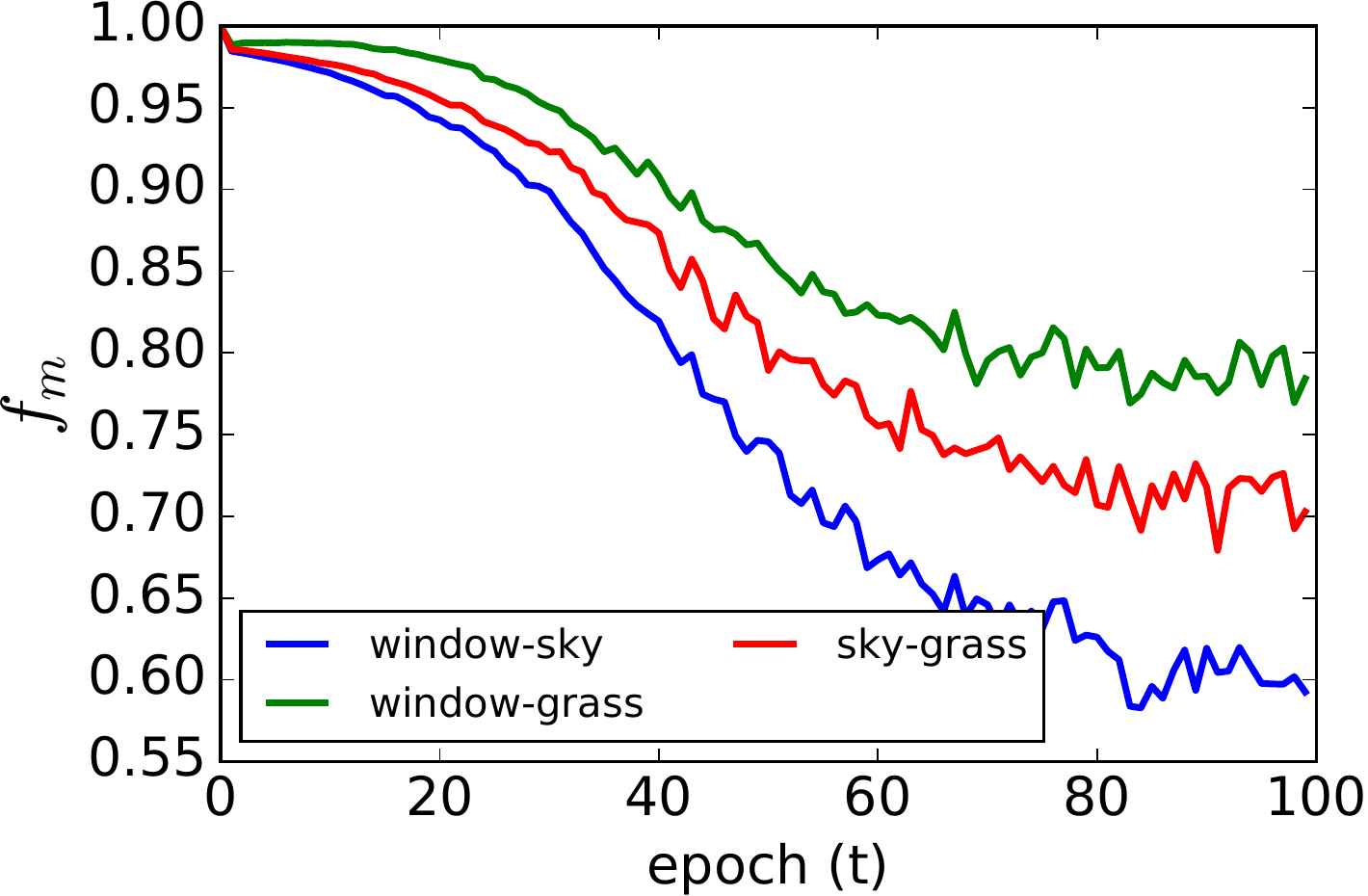} }}%
    \qquad
    \subfloat[t-SNE projections]{{\includegraphics[width=0.42\linewidth]{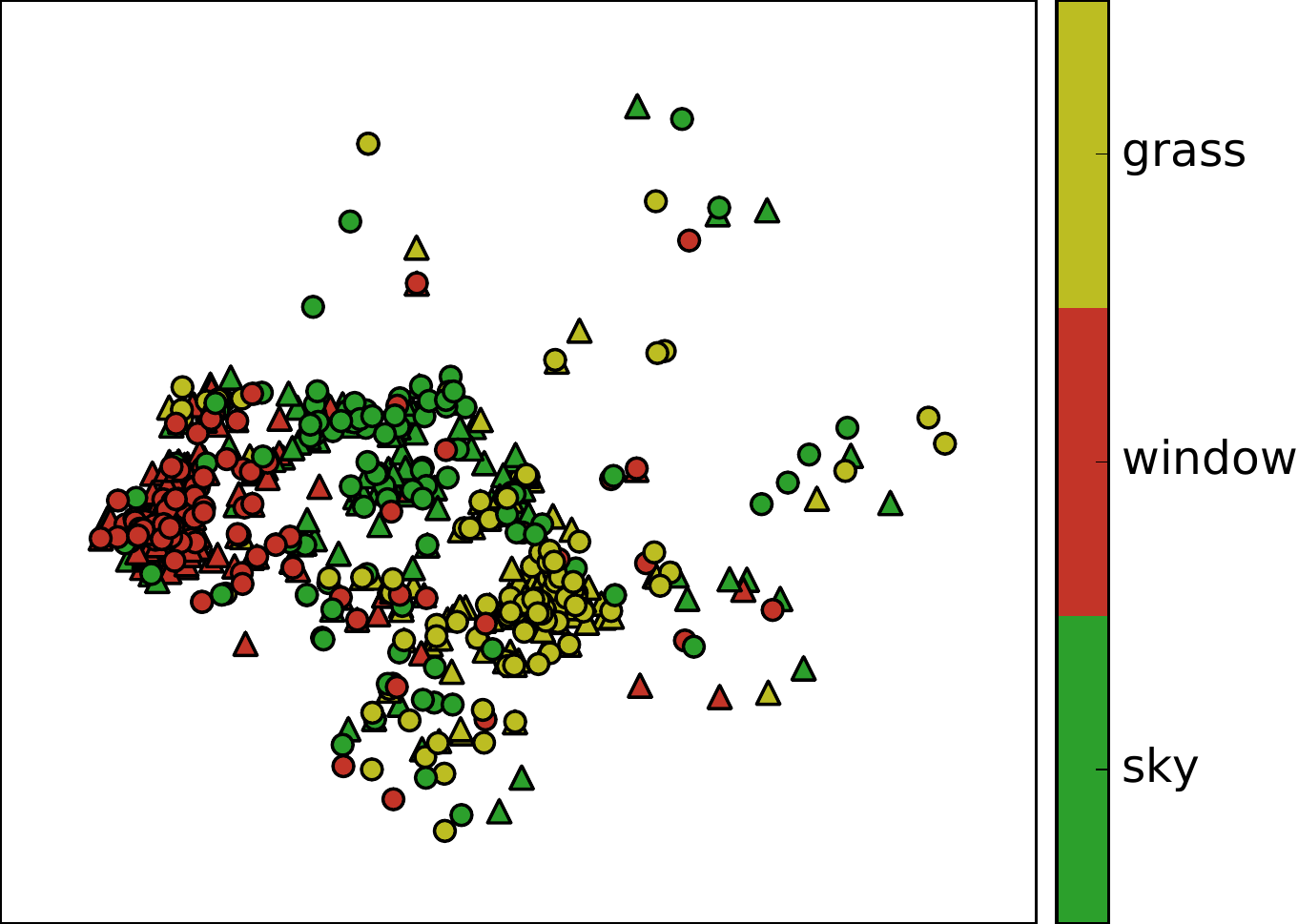} }}%
    \vspace{-3.5mm}
    \caption{Analysis of the margin values over each epoch (t), between three categories (left), versus final t-SNE model projections (right). }%
    \label{fig:margin_evolution_vs_projection}%
    \end{minipage}
    \vspace{-2mm}
\end{figure*}

\vspace{2mm}
\noindent
\textbf{Large-scale NUS-WIDE.}
To further explore the generalisation of SAM algorithm, we evaluated SAM in the large-scale full NUS-WIDE dataset. Table~\ref{table:nus-wide} supports the same conclusions that where drawn from the previous analysis. It is also noticeable, that all models improved thanks to the larger training dataset.

\vspace{5pt}
\noindent\textbf{Overview.} In overall, our method has proven to be effective, outperforming previous state-of-the-art methods on all datasets. The cluster enforcement and preservation component ($f_{mc}$) proved to be crucial to achieve state-of-the-art performance. Unlike most methods, which impose extra constraints by augmenting a projection network by adding additional loss terms, our approach imposes those constraints by directly adapting the margin between instance pairs during training, thus resulting in a simpler but effective model.

By modelling the semantic inter-category pairwise correlations, our model is able to transfer semantic correlations from the original feature space directly to the common subspace. Then, by enforcing cluster formation after achieving a stable subspace organisation, our method improves significantly from state-of-the-art works.

\begin{table}[t]
\caption{mAP results on the NUS-WIDE dataset.}
\label{table:nus-wide}
\vspace{-8pt}
\centering
\begin{tabular}{lccc}
\toprule
  \multirow{1}{*}{\textbf{}}&  \multicolumn{3}{c}{\textbf{NUS-WIDE}} \\
 \multirow{1}{*}{\textbf{Methods}} & $I\mapsto T$ & $T\mapsto I$ & Avg.\\
\hline
CCA~\cite{hotelling36cca} & 0.244 &0.275 &0.260 \\
CFA~\cite{Li:2003:MCP:957013.957143} &0.358 &0.361 &0.360\\
KCCA~\cite{6788402} &0.348 &0.481 &0.415\\
LGCFL~\cite{7006724} &0.512 &0.600 &0.556\\
JRL~\cite{6587747} &0.615 &0.592 &0.604\\
\hline
Corr-AE~\cite{Feng:2014:CRC:2647868.2654902} &0.391 &0.429 &0.410\\
DCCA~\cite{7298966} & 0.475 &0.500 &0.488\\
CMDN~\cite{Peng:2016:CSR:3061053.3061157}  &0.643 &0.626 &0.635 \\
CCL~\cite{8013822}& \underline{0.671} &  \underline{0.676} &  \underline{0.674} \\
\hline
\textbf{SAM} &  \textbf{0.701} & \textbf{0.707} & \textbf{0.704}\\
\bottomrule
\end{tabular}%
\vspace{-15pt}
\end{table}%

\subsection{Scheduled adaptive margins analysis}
In this section we examine behaviour of SAM by looking at the margin values imposed by the model on each triplet constraints, over each epoch (t).

\subsubsection{Average margin values vs. scheduler function}
The scheduler function $\alpha(t)$ shifts from a high-magnitude constant margin ($m=1$), to the adaptive margin $f_{am}$. To inspect this behaviour, we computed the average margin value, imposed to all triplets, on each epoch t, on the NUSWIDE-10k dataset. Figure~\ref{fig:margin_statistics_avg_epoch} shows the average $f_m$ value (blue line) versus the scheduler function value $\alpha(t)$ (green line), over the training epochs. It can be observed that at each epoch, the average margin imposed by $f_m$ tends to be smaller. One can also observe that $\alpha(t)$ has a sigmoidal shape.

\subsubsection{Average margin values for each Category}
In order to provide a deeper understanding of what the model achieves, we show in Figure~\ref{fig:margin_evolution_vs_projection}, also on the NUSWIDE-10k dataset, the average margin values between three pairs of categories at each training \textit{epoch t}, and a projection of the final cross-modal subspace.

The scale of the average margin values in the last epoch ($t=100$), between each pair of the considered categories, is reflected in the obtained subspace. It is noteworthy to say that the magnitude of the value $m$ reflects the difference between similarities of pairs of instances, not distance on the subspace. Nevertheless, the magnitude of the values still allow to confirm its impact in the subspace organisation. For instance, in the plot on the left, it can be seen that in the last epochs, our model enforced an average margin of roughly $0.6$ between instances of category \textit{window} versus category \textit{sky}, which is much smaller than the value between instances of \textit{window} and \textit{grass}, which is roughly $0.77$. Looking at the t-SNE projections, we can actually see that the organisation of instances respects these values, with instances of category \emph{window} being closer to instances of \emph{sky} than to \emph{grass}.

These experiments are crucial to understand the underpinnings of SAM: Figure~\ref{fig:margin_evolution_vs_projection} confirms that the average margin value gradually decreases during training, with triplet constraints over \emph{window}-\emph{sky} categories having lower magnitude margins than \emph{window}-\emph{grass}, thus reflecting visual and textual semantic similarity as intended.

\begin{figure}[!t]
    \centering
    \subfloat{\includegraphics[width=0.50\linewidth]{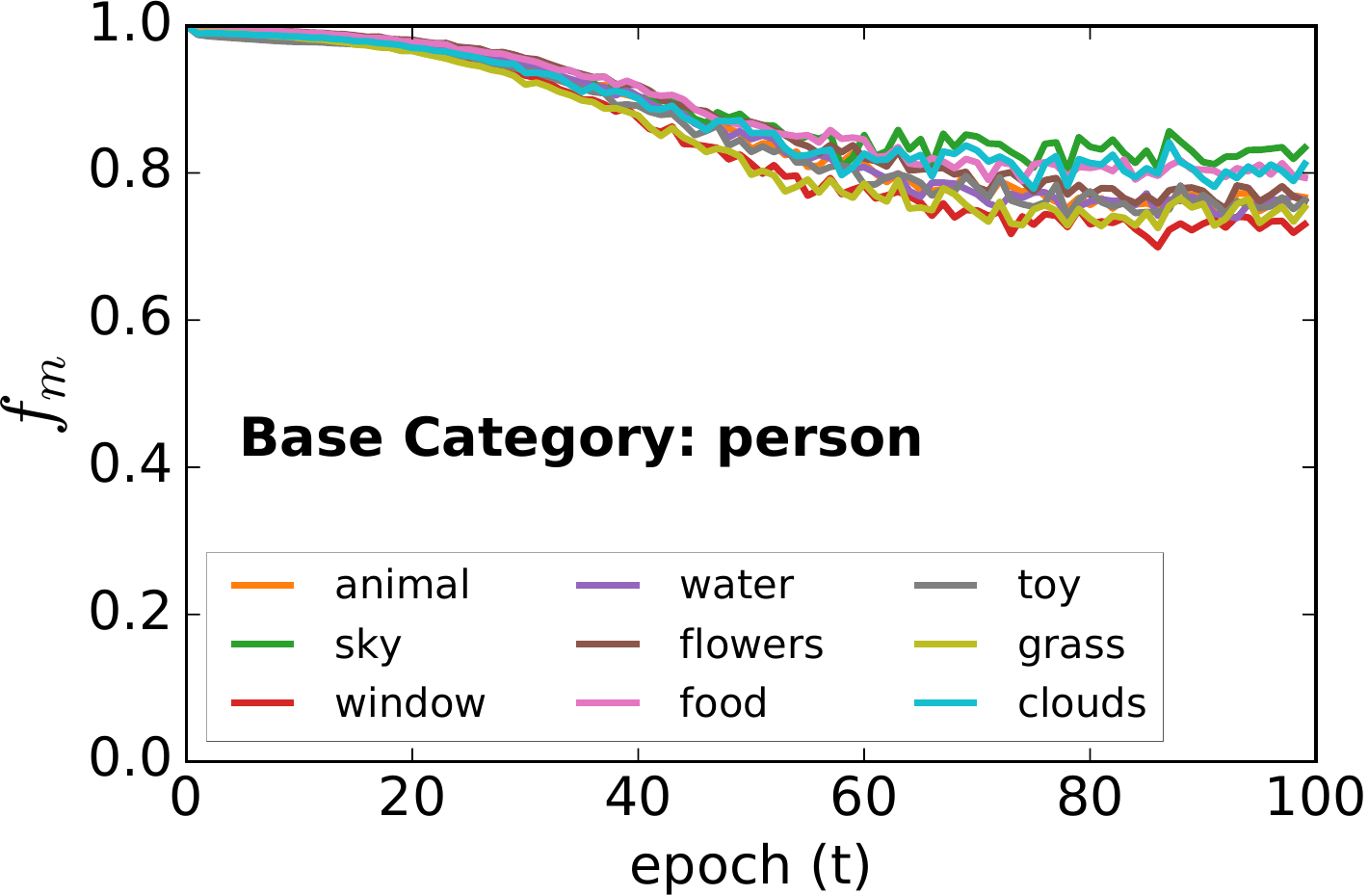}}
    \subfloat{\includegraphics[width=0.50\linewidth]{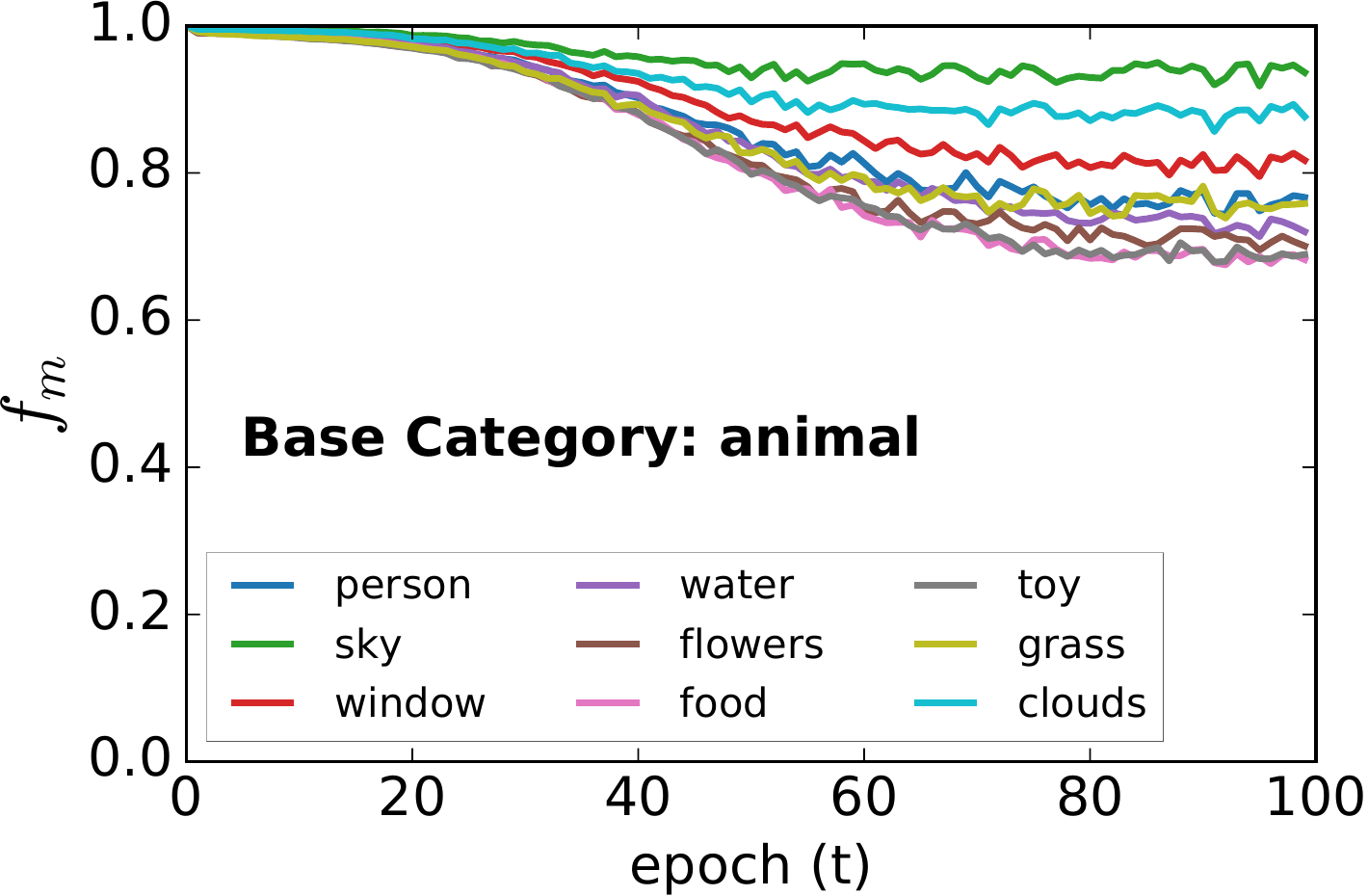}}   \\
    \subfloat{\includegraphics[width=0.50\linewidth]{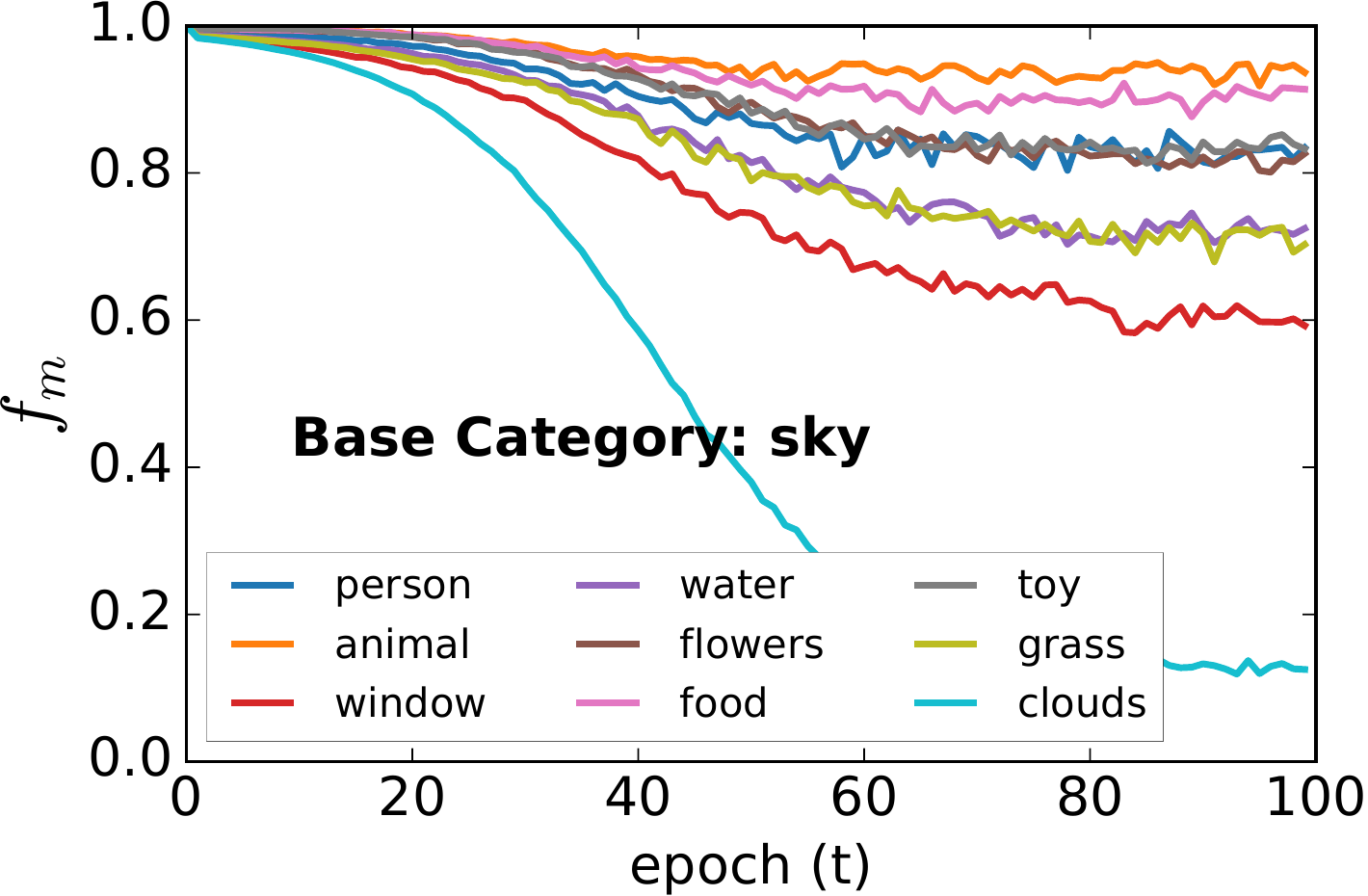}}
    \subfloat{\includegraphics[width=0.50\linewidth]{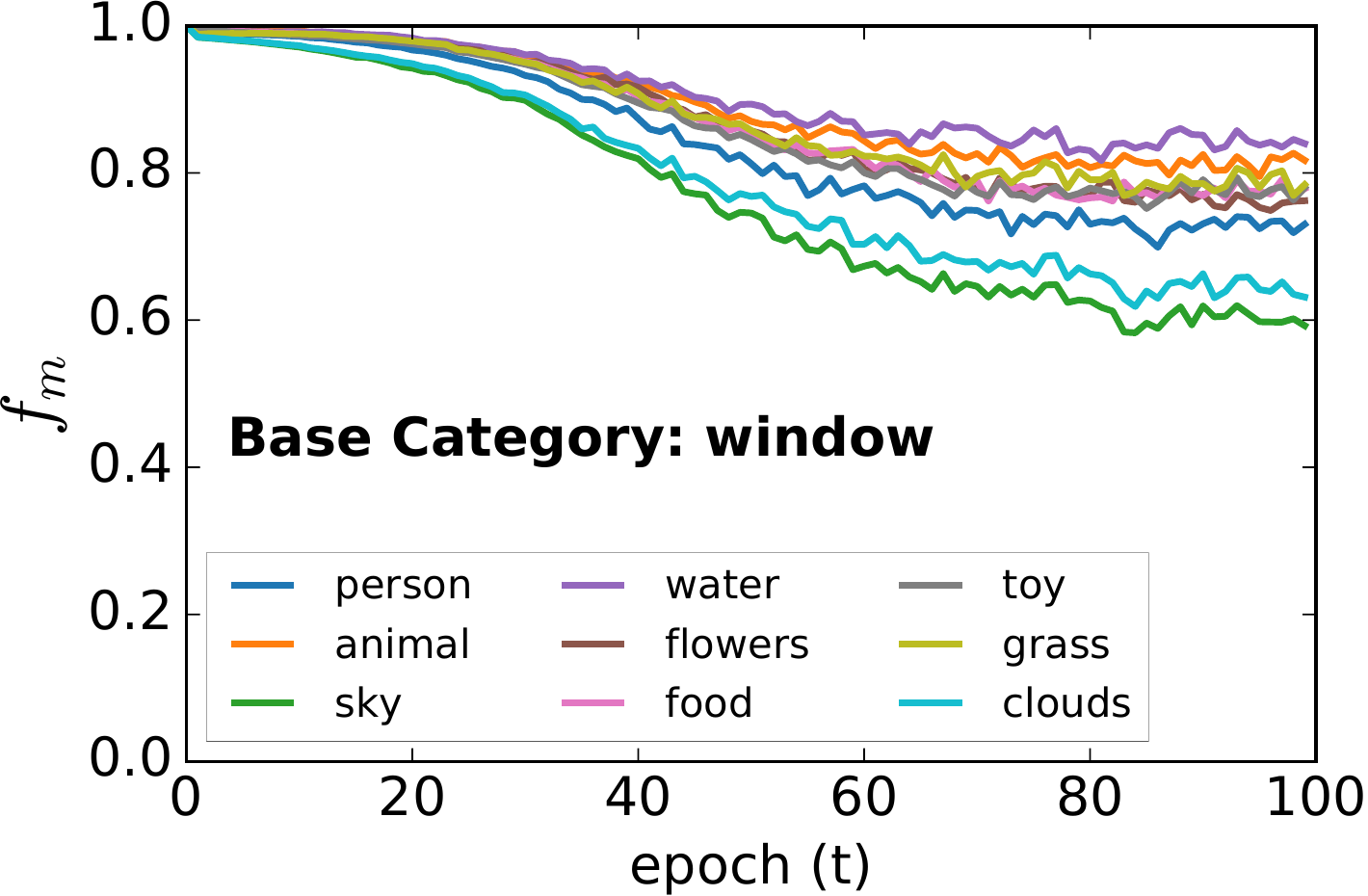}}    \\
    \subfloat{\includegraphics[width=0.50\linewidth]{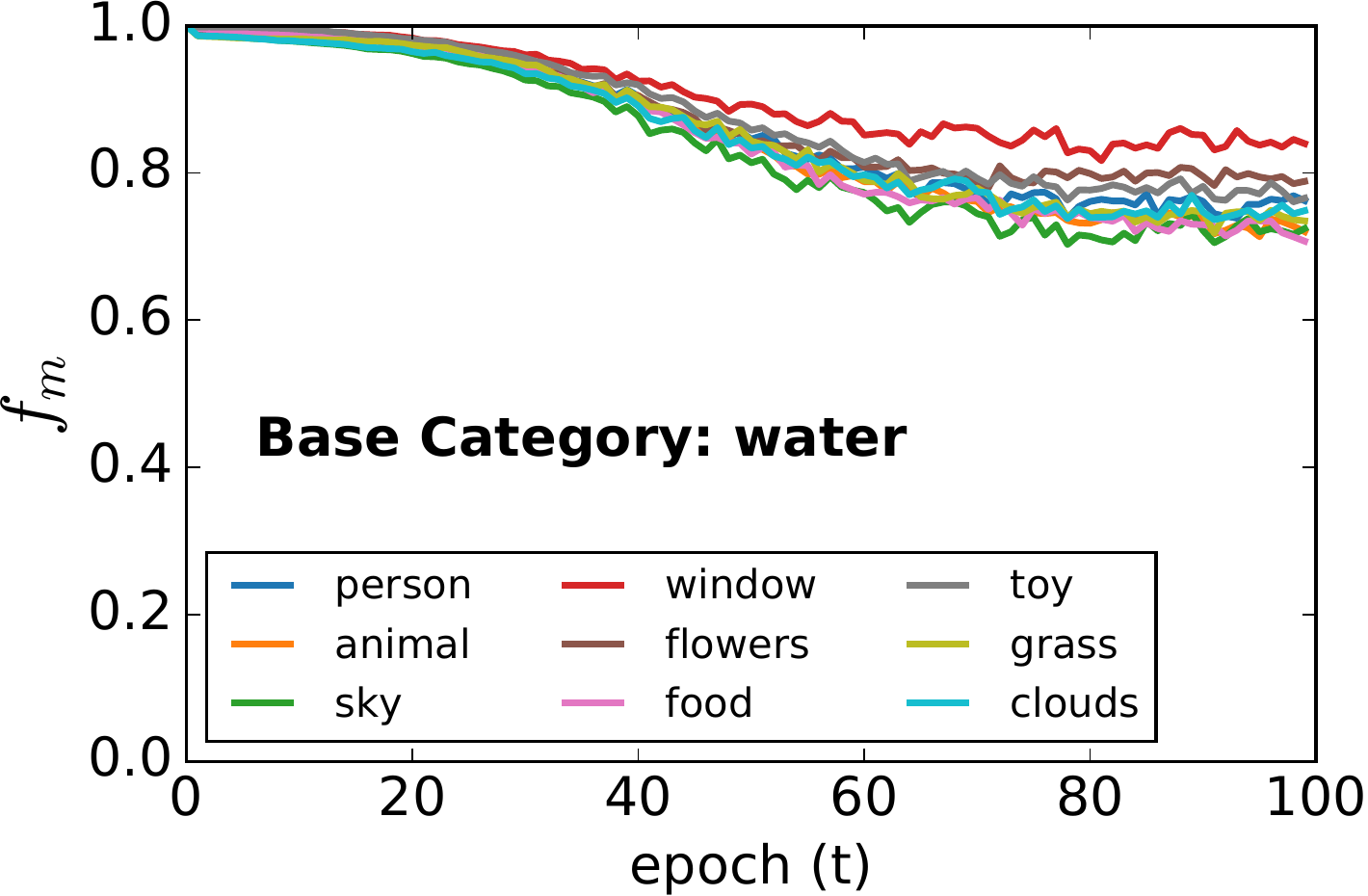}}
    \subfloat{\includegraphics[width=0.50\linewidth]{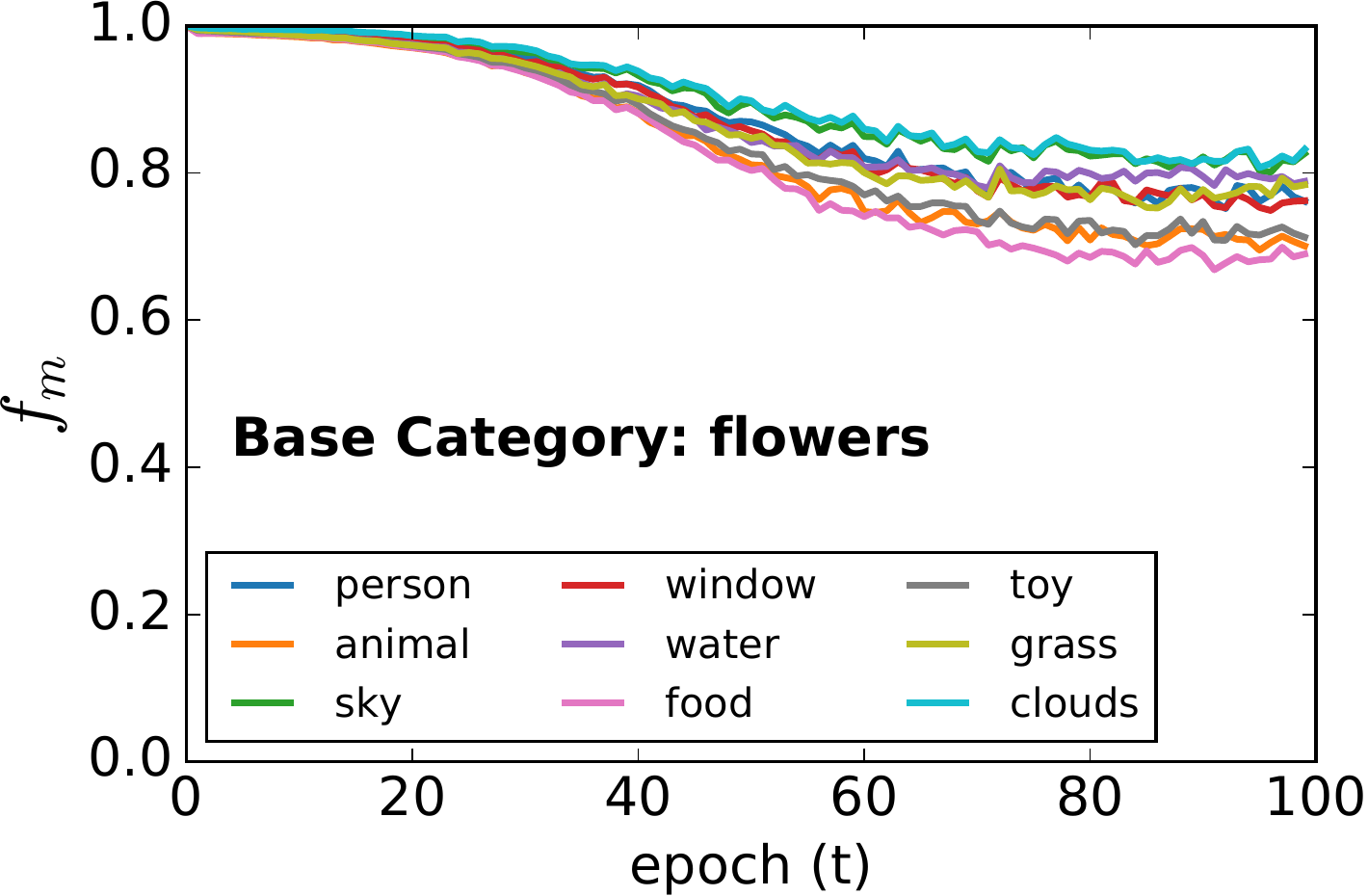}} \\
    \subfloat{\includegraphics[width=0.50\linewidth]{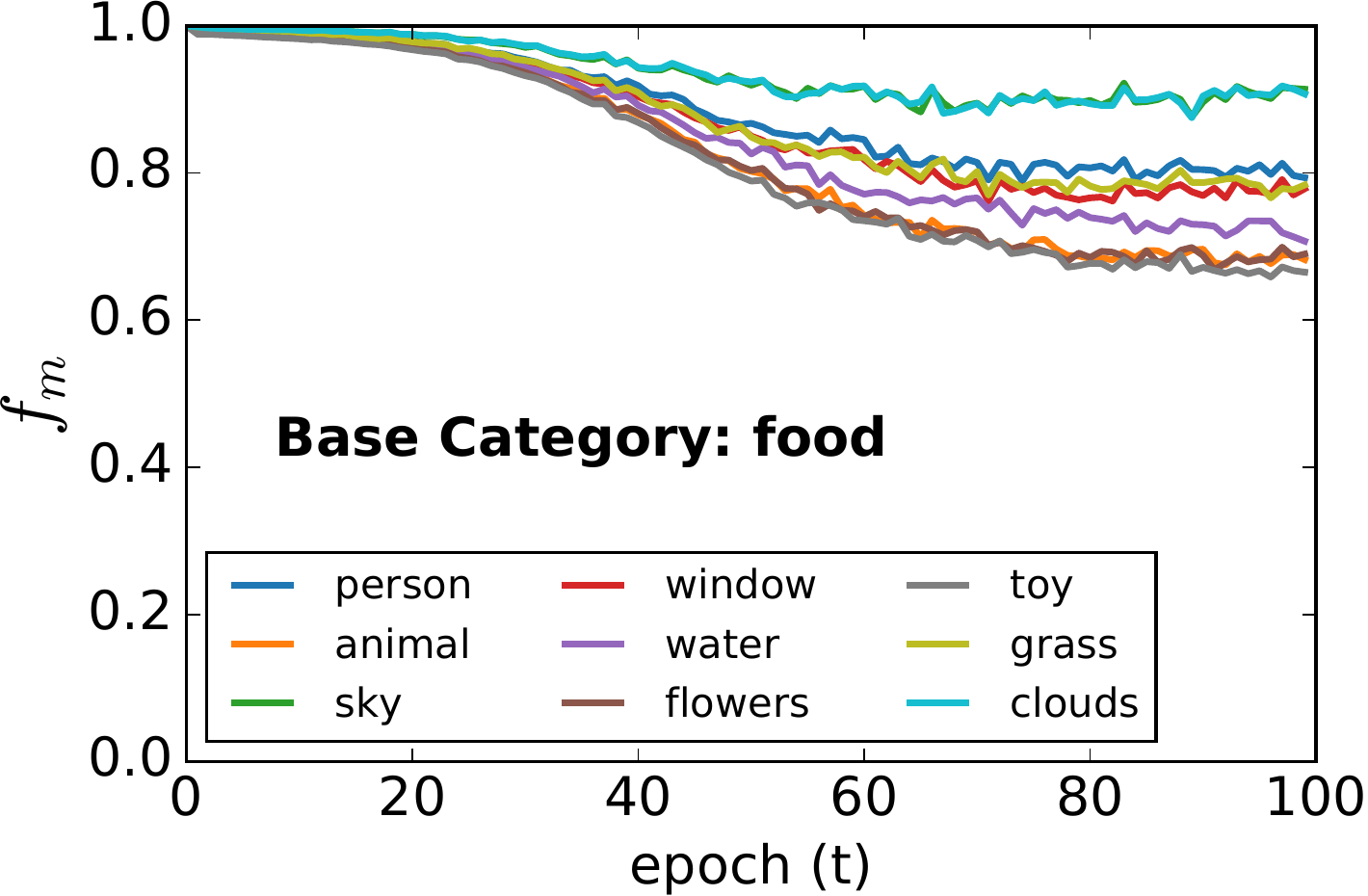}}
    \subfloat{\includegraphics[width=0.50\linewidth]{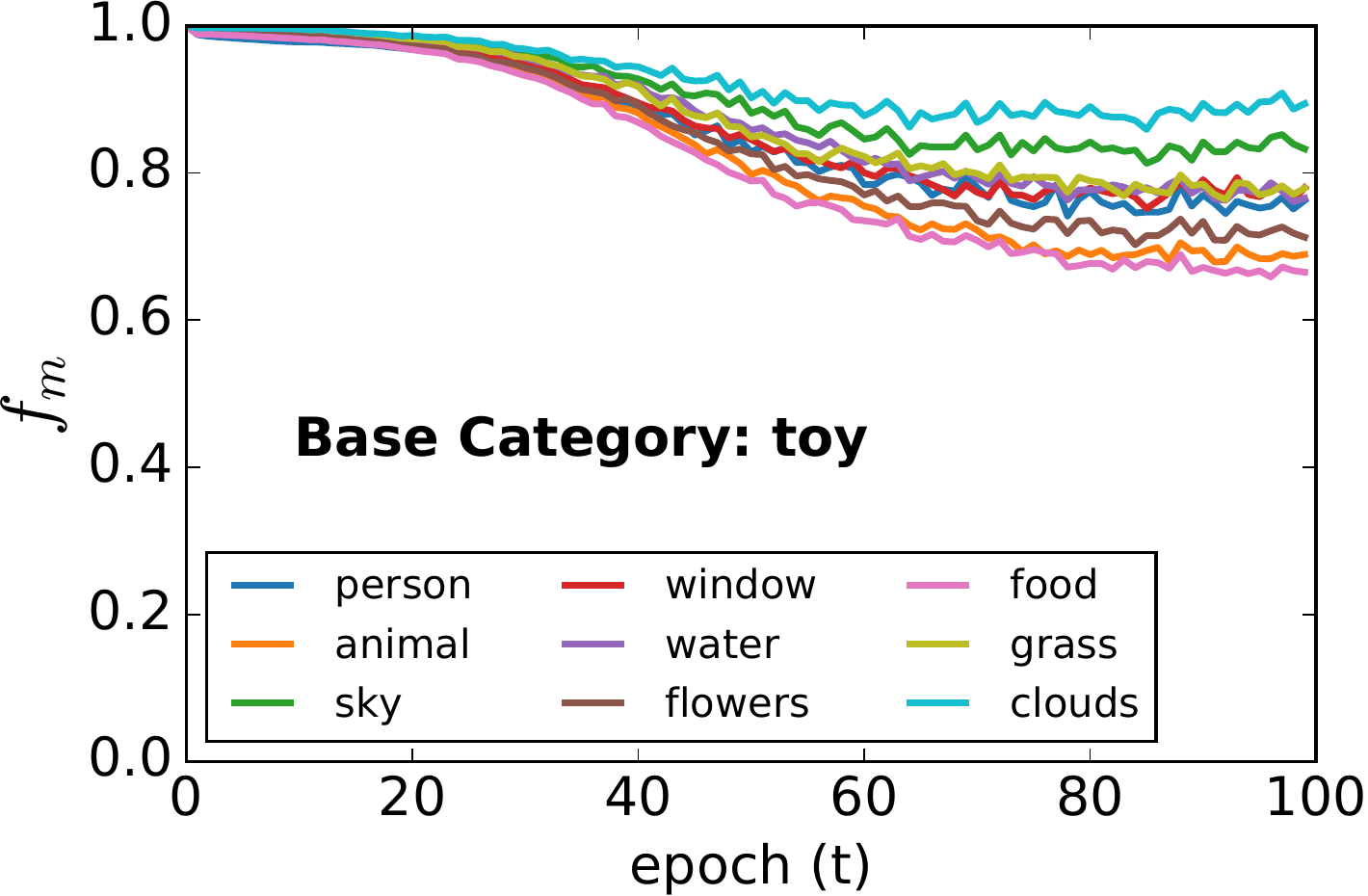}}   \\
    \subfloat{\includegraphics[width=0.50\linewidth]{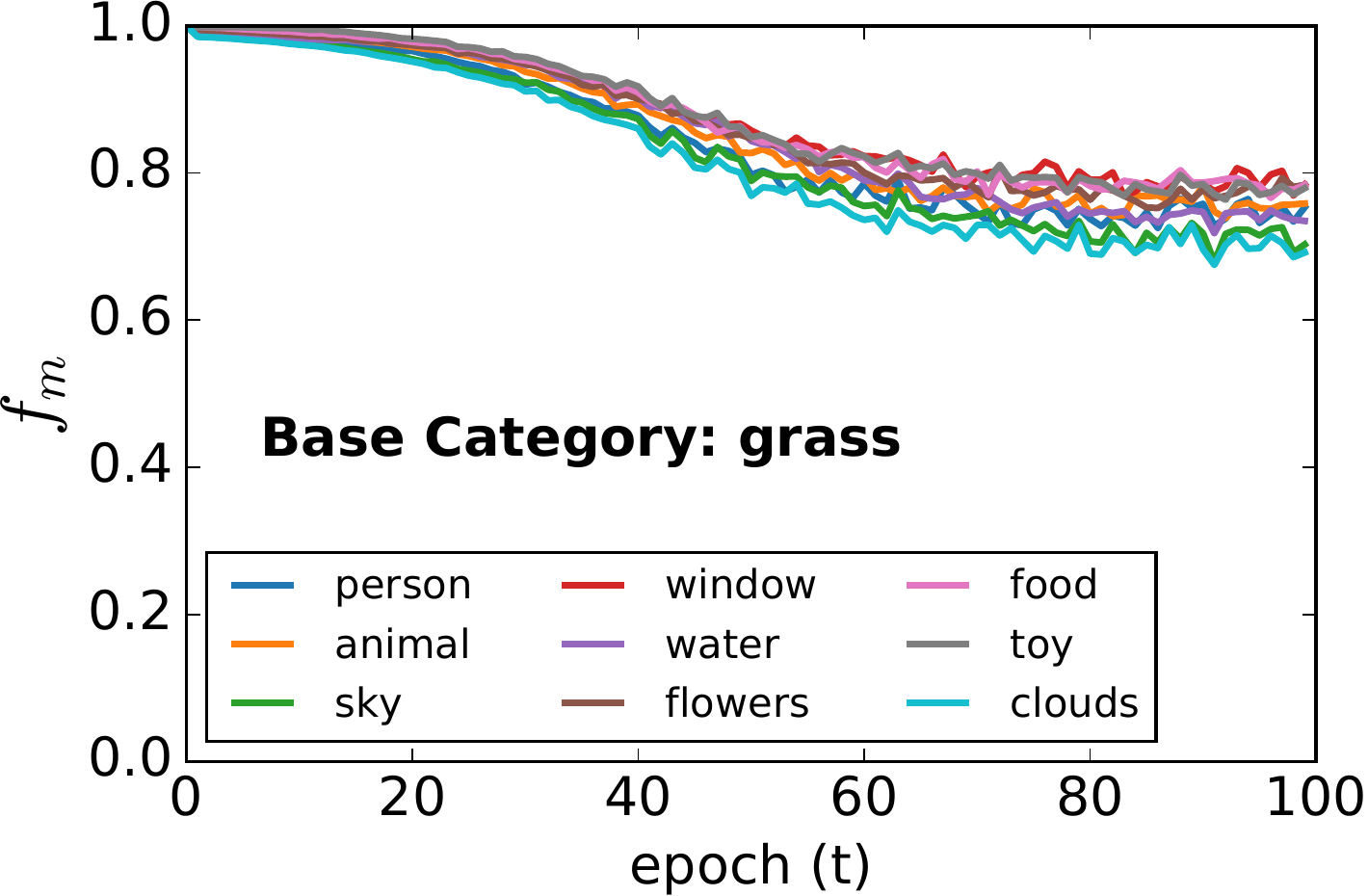}}
    \subfloat{\includegraphics[width=0.50\linewidth]{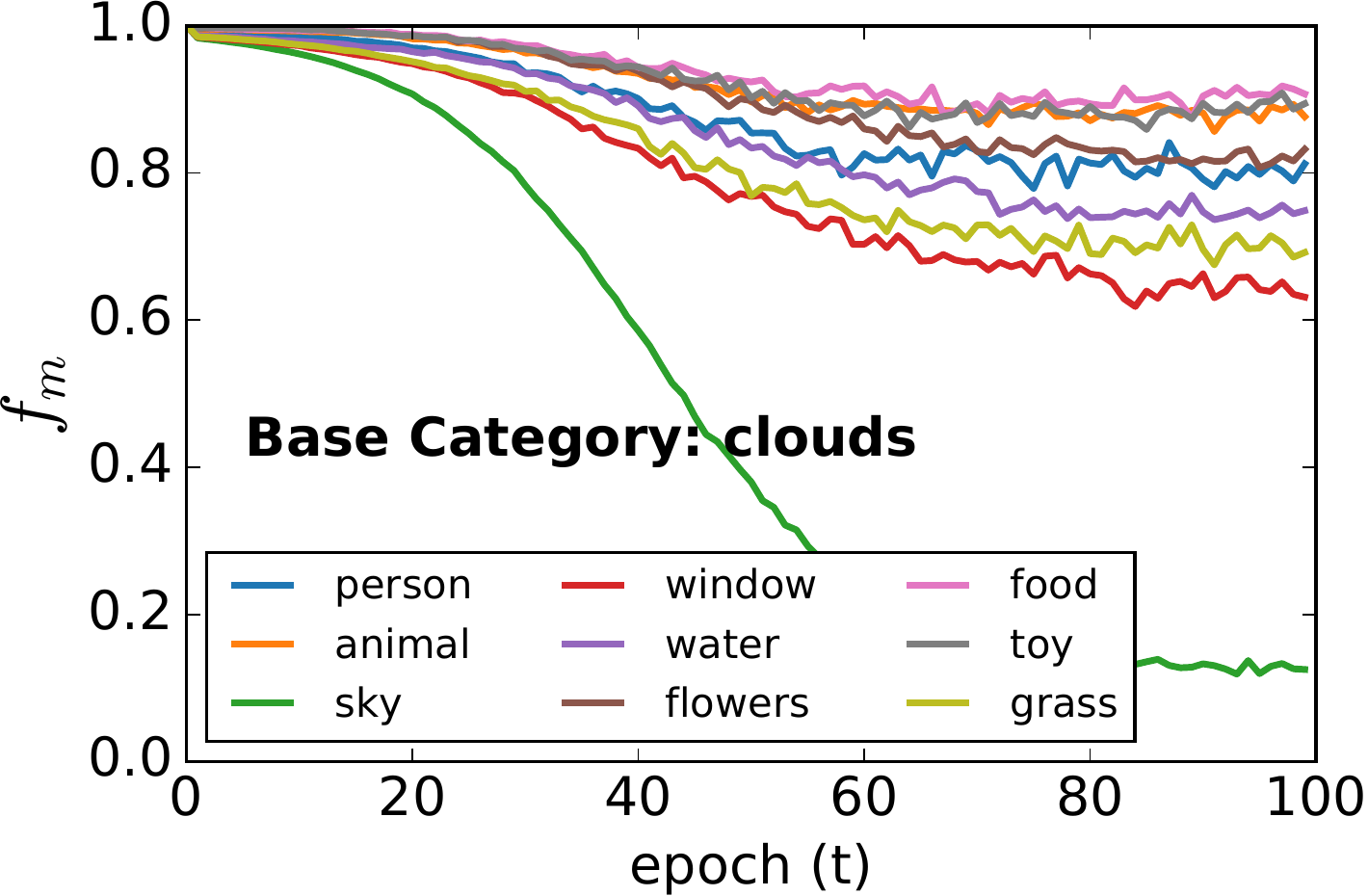}}
    \vspace{-4pt}
\caption{Average per-category margin for each category, at each training epoch (t). Average value of $f_m$ between every instance $d^i$, against all instances $d^n$ of other categories.}
  \label{fig:margin_per_category_9}
  \vspace{-5mm}
\end{figure}

Figure~\ref{fig:margin_per_category_9} delves into this question and shows the average margin value per category imposed by $f_m$, against triplets of the remaining categories, at each epoch $t$. Given the target category $c_1$ of each plot, each line corresponds to a category $c_2$. Namely, it corresponds to the average of the margin values, imposed by $f_m$, to triplets with the positive instance belonging to category $c_1$ and the negative belonging to category $c_2$.
It is interesting to note that all margins are significantly different. In particular, categories \textit{grass} and \textit{person} are the ones with most homogenous margins. In contrast, categories \textit{sky} and \textit{animal} took full advantage of the scheduled adaptive margins and ended up with very different margins to all other categories.

\subsubsection{Scheduler and $f_{mc}$ impact}
The scheduler, together with the cluster formation and enforcement $f_{mc}$ component of the adaptive margin, are key novel components, responsible for achieving state-of-the-art performance. To confirm this, we evaluated SAM with the scheduler deactivated ($\alpha(t) = 1)$ and with $f_{mc}$ disabled ($\lambda = 1$). As can been seen from table~\ref{table:results}, this results in a drop of performance of $\approx 8\%$, $\approx5\%$ and $\approx19\%$, on Pascal Sentences, NUS-WIDE-10k and Wikipedia, respectively, confirming the crucial importance of the scheduler and $f_{mc}$.

\vspace{-10pt}
\subsection{Analysis of activation phase}
\begin{figure}[t]
    \centering
    \includegraphics[width=0.75\linewidth]{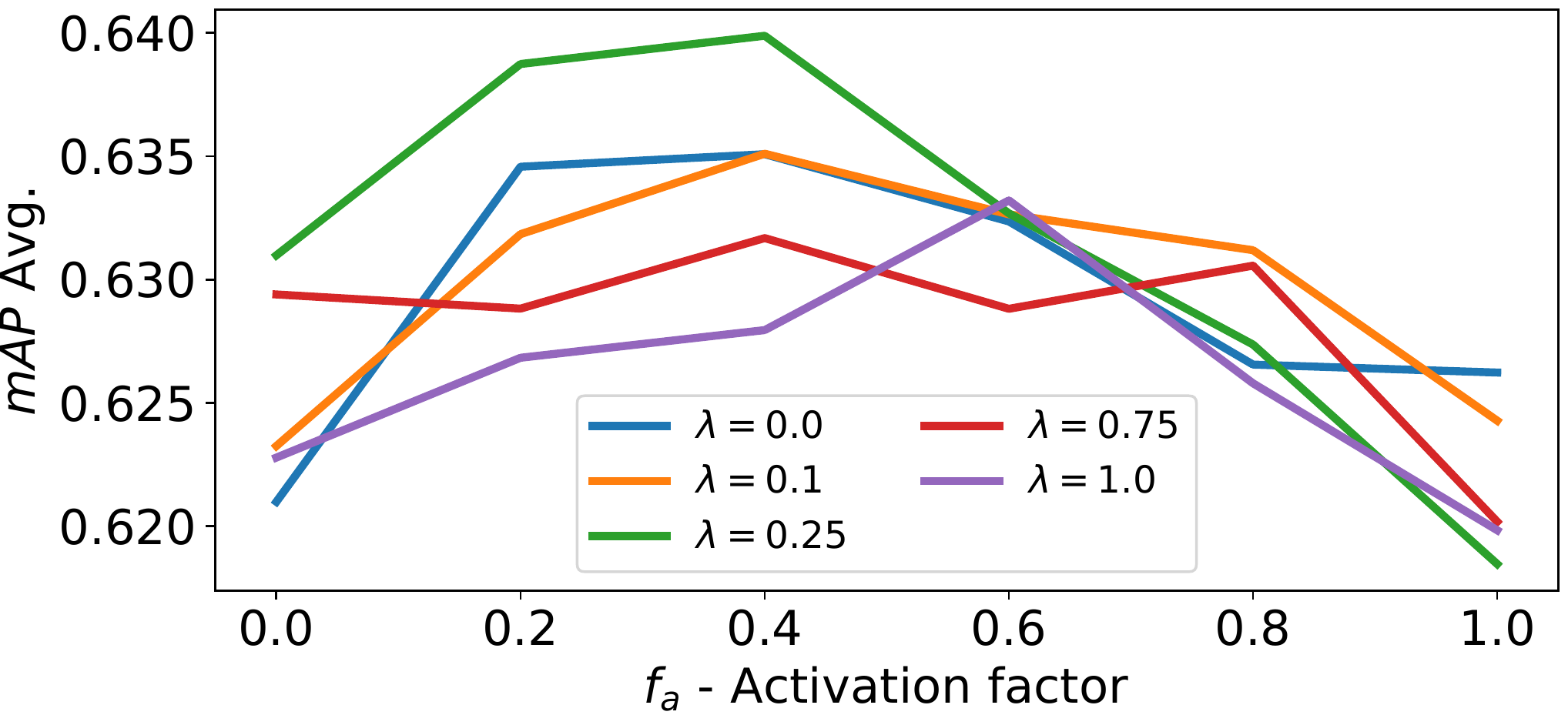}
\caption{Parameter Analysis ($\lambda$ and activation function $f_a$) on Pascal Sentences dataset. }
  \label{fig:parameters_analysis}
  \vspace{-15pt}
\end{figure}

In this section we will analyse the impact of the activation phase $f_a$ and the semantic correlation vs. cluster enforcement trade-off $\lambda$ parameter. To do this, we measure the $mAP$ score on the Pascal Sentences dataset. Namely, we evaluate $f_a \in \{0.0., 0.2, 0.4, 0.6, 0.8, 1.0 \}$ and $\lambda \in \{0.0, 0.1, 0.25, 0.75, 1.0\}$, fixing all the remaining parameters, and show the results in Figure~\ref{fig:parameters_analysis}. The $x$-axis represents the value of $f_a$ and the $y$-axis the $mAP$ score obtained. Each curve corresponds to a value of $\lambda$.

The first observation is that imposing the adaptive margin too early is bad. For instance, when $f_a$ is close to zero, the method starts using the adaptive margin from the beginning of the training, resulting in low performance. This confirms our intuition that in the first training iterations, the subspace is still coarsely organised. As the parameter $f_a$ increases, we can see that the results improve significantly, reaching a performance peak on $f_a=0.4$, for four of the five experimented values of $\lambda$. Namely, smoothly activating the adaptive margin with $f_a=0.4$, and giving around $75\%$ weight to $f_{mc}$ (cluster formation and preservation component) and the remaining to $f_{ms}$, leads to the best performance.
For all values of $\lambda$, activating the adaptive margin too late leads to significant performance drops. This is due to the fact that by activating later, the network has more chances to overfit using a static margin. At this point, neither the cluster formation $f_{mc}$, nor the semantic correlations $f_{ms}$ components are able to improve the subspace organisation.
Regarding the trade-off parameter $\lambda$, we observe the trend that cluster formation has a higher impact on achieving better performance than semantic correlation, with peak performance occurring when both components are active.

\section{Conclusions}
In this paper we described a novel method to learn cross-modal embeddings. The method introduces a scheduled activation of adaptive margins that allow for triplet specific margins. The key takeaways of the proposed method are:
\begin{itemize}
    \item \textbf{Adaptive margin constraints:} our approach impose general constraints while training the model by adapting the margins between instance pairs. This overcomes the fact that using a unique margin for all pairs is insufficient to adequately structure the subspace.
    \item \textbf{Effective learning of pair-specific margins:} results show that adaptive margins deliver state-of-the-art results. This is further possible due to the pair-specific margins that are learned by the model as illustrated by experimental results.
    \item \textbf{Scheduled learning:}  new neural-network training approach was introduced that progressively activates the adaptive margin function, through an epoch-aware scheduling strategy.
\end{itemize}
As future work, we plan to generalise adaptive ranking loss. Current results hint that the same principle can encode different constraints and thus be extended to other multimedia modelling tasks.

\vspace{-5pt}
\begin{acks}
This work has been partially funded by the \grantsponsor{CMUP}{CMU Portugal}{} research project GoLocal Ref. \grantnum{CMUP}{CMUP-ERI/TIC/0046/2014}, by the \grantsponsor{EU H2020}{H2020 ICT}{} project COGNITUS with the grant agreement n\textsuperscript{o} \grantnum{EU H2020}{687605} and by the \grantsponsor{FCT}{FCT}{} project NOVA LINCS Ref. \grantnum{FCT}{UID/CEC/04516/2019}. We also gratefully acknowledge the support of NVIDIA Corporation with the donation of the GPUs used for this research.
\end{acks}

\balance
\bibliographystyle{ACM-Reference-Format}
\bibliography{01_bibliography}

\end{document}